\documentclass[english,letterpaper,english,prb,notitlepage,twocolumn,superscriptaddress,floatfix]{revtex4-2}
\usepackage[utf8]{inputenc}
\usepackage{amsmath, amssymb, dsfont, cancel, nicefrac, braket}
\usepackage{graphicx, float, subfiles, stackengine}
\usepackage[caption=false]{subfig}
\usepackage{hyperref, color, ulem, soul}
\usepackage{booktabs}
\usepackage[usenames,dvipsnames,svgnames,table]{xcolor}
\usepackage{comment}

\definecolor{violet}{rgb}{0.58, 0.0, 0.83}

\newcommand{\mat}[1]{\pmb{#1}}

\newcommand{\Tr}{\operatorname{Tr}}
\newcommand{\hc}{\text{h.c.}}

\newcommand{\op}[1]{\hat{#1}}
\newcommand{\supop}[1]{\mathbb{#1}}

\newcommand{\stkout}[1]{\ifmmode\text{\sout{\ensuremath{#1}}}\else\sout{#1}\fi}
\newcommand{\nnref}[3]{\hyperref[#1]{#2\ref*{#1}#3}}

\newcommand{\bigO}{\mathcal{O}}

\newcommand{\dop}[3]{\,\Big\{ |#1\rangle\langle #2| \Big\} \,}

\newcommand{\gF}[2]{{#1}^{+}_{#2}}
\newcommand{\gFc}[2]{{\overline{#1}}^{+}_{#2}}
\newcommand{\gB}[2]{{#1}^{-}_{#2}}
\newcommand{\gBc}[2]{{\overline{#1}}^{-}_{#2}}
\newcommand{\roi}[2]{{\color{red} \op{1}(#1, #2 )}}

\newcommand{\ubar}[1]{\mkern 1.5mu\underline{\mkern-1.5mu#1\mkern-1.5mu}\mkern 1.5mu}

\newcommand{\nnaddcomment}[1]{}
\newcommand{\nndel}[1]{}
\newcommand{\nndelmath}[1]{}

\makeatletter
\let\inserttitle\@title
\makeatother

\begin{document}

\title{Real time evolution of Anderson impurity models via tensor network influence functionals}
\author{Nathan Ng}
\affiliation{Department of Chemistry, Columbia University, New York, New York 10027, United States}
\author{Gunhee Park}
\affiliation{Division of Engineering and Applied Science, California Institute of Technology, Pasadena, CA 91125, USA}
\author{Andrew J.\ Millis}
\affiliation{Department of Physics, Columbia University, New York, New York 10027, USA}
\affiliation{Center for Computational Quantum Physics, Flatiron Institute, New York, New York 10010, USA}
\author{Garnet Kin-Lic Chan}
\affiliation{Division of Chemistry and Chemical Engineering, California Institute of Technology, Pasadena, California 91125, USA}
\author{David R.\ Reichman}
\affiliation{Department of Chemistry, Columbia University, New York, New York 10027, United States}
\makeatletter
\let\inserttitle\@title
\makeatother

\date{\today}

\begin{abstract}
In this work we present and analyze two tensor network-based influence functional approaches for simulating the real-time dynamics of quantum impurity models such as the Anderson model. Via comparison with recent numerically exact simulations, we show that such methods accurately capture the long-time non-equilibrium quench dynamics. 
The two parameters that must be controlled in these tensor network influence functional approaches are a time discretization (Trotter) error and a bond dimension (tensor network truncation) error. We show that the actual numerical uncertainties are controlled by an intricate interplay of these two approximations which we demonstrate in different regimes. Our work opens the door to using these tensor network influence functional methods as general impurity solvers.
\end{abstract}

\maketitle

\section{Introduction}

The numerical simulation of quantum many-body systems is in principle an exponentially difficult problem due to growth of Hilbert space dimension with system size.
While the advent of tensor network techniques has made the equilibrium problem more tractable, the  application of tensor network methods to dynamics has been limited by the growth of entanglement during time evolution, which can cause exponentially growing resource requirements to accurately describe the dynamics.
Given the complex nature of this problem, developing new numerical methods and furthering understanding of dynamical phenomena is best done within the context of simple, yet nontrivial models.

Quantum impurity models offer one such possibility, with the additional benefit that they are of practical physical importance, having  led to an understanding of phenomena ranging from the Kondo effect in solids~\cite{Hewson1993} to the survival of macroscopic quantum coherence effects in condensed phases~\cite{Leggett1987, Weiss2021}. 
In the past few decades, impurity models have gained additional significance as a key component of computational embedding frameworks such as dynamical mean field theory~\cite{Georges1996} and density matrix embedding theory~\cite{Knizia2012, Kretchmer2018}.  Here, one must compute the real-time dynamics of the impurity to obtain an approximation to the local dynamics of the full problem via a self-consistently refined effective bath.  Thus the efficacy of such embedding methods is constrained by the flexibility, speed, and accuracy with which the dynamics of an impurity coupled to a bath can be simulated.

Numerical approaches for solving impurity problems must contend with issues around the treatment of a large (infinite) number of bath degrees of freedom, while simultaneously needing to ameliorate the dynamical sign problem for impurity dynamics.
The issue of treating continuous baths poses an impediment to methods like diagonalization-based techniques~\cite{Liebsch2011, Granath2012, Lu2014, deVega2015}, and is partially solved by the advent of tensor network approaches~\cite{Guettge2013, Wolf2014, Ganahl2015, Bauernfeind2017, Bauernfeind2019, Kohn2022}. The treatment of the sign problem saddles dynamical Monte Carlo methods~\cite{Muehlbacher2008, Cohen2011, Gull2011, Gull2010, Cohen2013, Cohen2014a, Cohen2014b} with an exponentially scaling numerical cost. Although this scaling can be tamed by inchworm diagrammatic expansions~\cite{Cohen2015, Boag2018, Cai2020, Li2022}, these approaches still incur statistical errors. Methods based on the Feynman-Vernon influence functional (IF), such as iterative path integral methods~\cite{Makri1992, Makri1995a, Mundinar2022} and the hierarchical equations of motion (HEOM) approach~\cite{Tanimura1989, Ishizaki2005} have been used with great success in spin-boson-type problems but are far less explored in problems with fermionic baths such as 
the Anderson model~\footnote{See, however~\cite{Haertle2013, Haertle2015, Jin2008, Zheng2009, Dong2014, Li2012, Cirio2022}}.
Generally speaking, treating the problem of impurity dynamics using the IF provides a natural formalism to consider continuous baths without invoking stochastic sampling which introduces the dynamical sign problems.

Of the above approaches, discretized path integral-based IF methods, e.g.\ the quasi-adiabatic path integral method (QUAPI)~\cite{Makri1992, Makri1995a, Makri2017, Makri2020}, are promising, as they reduce the problem to one of managing Trotter errors and memory truncation.  However, models such as the Anderson model are more complex than the spin-boson model, in part because the system part of the system-bath coupling cannot be written in a simple diagonal form.  Early attempts to generalize these approaches to the dynamics of Anderson-like models provided a window to obtain exact non-equilibrium dynamics in some parameter regimes, but suffered from memory length issues that limited their range of applicability~\cite{Weiss2008, Segal2010}.  In the intervening years, significant progress has been made marrying tensor network methods and IF methods, particularly for spin-boson-like models.
Specifically, one can view the Trotterized dynamics as a tensor network and employ tensor contraction over the environment degrees of freedom for a fixed propagation time, instead of contracting along the temporal direction~\cite{Banuls2009, Hastings2015, Tirrito2018, Ye2021, Bose2021, Lerose2021}.
The resulting object can be viewed as a matrix product state representation of the exact discretized influence functional (MPS-IF), which exists on a temporal lattice.
Such an approach has been successfully applied to harmonic baths~\cite{Strathearn2018, Jorgensen2019, Ye2021, Bose2021, Bose2022a, Bose2022b, Gribben2022} as well as finite baths of bosons, fermions, or spins~\cite{Ye2021, Cygorek2022}.

Recently, Abanin and coworkers have focused on the behavior of MPS-IF approaches for fermionic models, and have usefully detailed how entanglement properties of the influence functional temporally evolve~\cite{Thoenniss2022, SONNER2021168677, Lerose2021}.
In this paper we follow Abanin et al, considering the explicit non-equilibrium dynamics of the Anderson impurity model directly in the continuous fermionic bath limit. We use a more general and flexible formulation of the system-bath coupling to allow for the treatment different bath densities of state.
In addition, we take advantage of the Gaussian nature of the bath to construct the MPS-IF in two ways, either by leveraging the Gaussian form of the IF when the bath is noninteracting, or by propagating it forward iteratively in a similar spirit to the QUAPI method~\cite{Makri1995a}. Using this approach we present a  numerical solution of the Anderson impurity model including a comparative analysis and optimization  of the errors arising from the required time discretization and bond dimension truncation.  We thus obtain approaches to the impurity problem where the convergence is not determined by  the standard bath size and sign issues, and which can produce a description of the true non-Markovian evolution of the impurity with polynomially-scaling numerical effort and error control. 

This paper is organized as follows: In Sec.~II we briefly outline the approach to real time dynamics using discretized influence functionals and we give two schemes for constructing a matrix product state representation for the IF.  In Sec.~III we compare our approach for the non-equilibrium dynamics of the Anderson model to recent exact calculations on the model. This comparison points to some important specific details of our approaches with respect to convergence which are then discussed in detail.  In Sec.~IV we conclude and discuss outstanding questions for future study.  Details of derivations are contained in the Supplementary Information.

\section{Influence functionals and their representations}
We consider the quench dynamics of the single impurity Anderson model,
\begin{align}
\label{eq:SIAM-hamiltonian}
    \begin{split}
        \op{H} &=  \sum_{k, \sigma} E_{k, \sigma} \op{c}^\dagger_{k, \sigma} \op{c}_{k,\sigma} + \sum_{k, \sigma} \Big( V_{k,\sigma} \op{c}^\dagger_{k,\sigma} \op{d}_{\sigma} + \hc \Big) \\
        &\phantom{=} + U \op{n}_{\uparrow} \op{n}_{\downarrow} + \sum_{\sigma} \varepsilon_{\sigma} \op{n}_{\sigma}.
    \end{split}
\end{align}
Here, $\varepsilon_{\sigma}$ is the on-site energy for electrons with spin $\sigma = \{\uparrow, \downarrow\}$ residing on the impurity,  $U$ is the Coulomb repulsion for two electrons that reside on the impurity, $E_{k, \sigma}$ is the conduction (bath) electron energy with momentum $k$, and $V_{k, \sigma}$ characterizes the strength of the coupling between the impurity and bath electrons. 
For the remainder of this paper we will refer to the terms in the first and second lines of \nnref{eq:SIAM-hamiltonian}{Eq.~(}{)} as $\op{H}_0$ and $\op{H}_1$ respectively; the bath-only terms within $\op{H}_0$ will also be denoted by $\op{H}_B$.

The dynamics we consider starts from an initially nonequilibrium state, in which the impurity is decoupled from the bath, $\op{\rho}_{\text{full}} = \op{\rho}(0) \op{\rho}_B$~\footnote{We note that the types of correlated initial conditions relevant for the calculation of equilibrium correlation (Green's) functions can be straightforwardly generated from imaginary time evolutions within the same formalism~\cite{Shao2002}.}.
The ensuing evolution of the impurity is approximated by a second-order Trotter decomposition as
\begin{align*}
    \begin{split}
        \op{\rho}(N \Delta t) &= \Tr_B \Bigg[ \Big( e^{-i \supop{L}_0 \frac{\Delta t}{2}} e^{-i \supop{L}_1 \Delta t} e^{-i \supop{L}_0 \frac{\Delta t}{2}} \Big)^{N} \Big\{ \op{\rho}(0) \op{\rho}_B \Big\}
        \Bigg]
    \end{split},
\end{align*}
where, for compactness, we have defined the superoperators $e^{-i \supop{L}_{0/1}} \op{A} \equiv e^{-i \op{H}_{0/1}} \op{A} e^{i \op{H}_{0/1}}$.
The trace over the bath can be performed in the basis of coherent states when the statistics of the bath is Gaussian, e.g. $\op{\rho}_B \propto \exp(-\beta \op{H}_B)$~\footnote{See Supplementary Materials at [] for additional data, along with details of the derivation of the IF and the computation using the MPS-IF.}.
What remains then is the impurity dynamics captured by trajectories over coherent states on the forward (backward) contour, $|\eta_n\rangle$ ($|\bar{\eta}_n\rangle$).
Each of these trajectories is weighted by the influence functional $I_N$ containing properties of the bath as well as the impurity-bath coupling, leading to the representation of the Trotterized dynamics as
\begin{align}
\label{eq:rdm-path-integral}
    \langle \eta^{*}_{N} | \op{\rho}(N \Delta t) | \bar{\eta}_{N} \rangle &= \int \Big( \prod_n \mathcal{D}\eta_n \mathcal{D}\bar{\eta}_n \Big) \,  I_N[\{ \eta_n \}, \{ \bar{\eta}_n \}] \\
    &\quad \times \langle \eta^{*}_0 | \op{\rho}(t_i) | \bar{\eta}_0 \rangle e^{i \int d\tau \, S_{\text{imp}}[\eta(\tau), \eta^{*}(\tau)]}. \nonumber
\end{align}
Since $\op{H}_0$ is quadratic by construction the influence functional takes the general Gaussian form,
\begin{align}
\label{eq:IF-quadratic-grassmann}
    I_N[\eta, \bar{\eta}] &= \exp \Bigg[ \begin{pmatrix} \bar{\eta}_1\\ \bar{\eta}_2 \\ \vdots   \end{pmatrix} \cdot \mat{G} \cdot \begin{pmatrix} \eta_1 & \eta_2 & \cdots \end{pmatrix},
    \Bigg],
\end{align}
where $\mat{G}$ is a matrix describing the temporal correlations in the impurity's trajectories.
For the single impurity case we consider here, the trajectory at time step $N$ is specified using four states. In particular, for each branch on the Keldysh contour (forward and backward), we must keep track of the impurity's state before and after it is acted on by $\exp(\pm i \op{H}_1 \Delta t)$ at time $N\Delta t$.
Thus, the IF at the $N$th time step requires $\mat{G}$ to be a $4N \times 4N$ matrix.
The explicit expressions of \nnref{eq:rdm-path-integral}{Eqs.~(}{)} and \nnref{eq:IF-quadratic-grassmann}{(}{)} are rather involved and are presented fully in~\cite{Note2}.

To represent and compute quantities associated with $I_N[\eta, \bar{\eta}]$, we can treat the Grassmann variables $\eta_n$ and $\bar{\eta}_n$ as fermionic operators~\cite{Gu2010, Thoenniss2022}.
This turns the $I_N[\eta, \bar{\eta}]$ into a generalized Gaussian state $|I_N\rangle$,
\begin{align}
\label{eq:IF-state-definition}
    |I_N\rangle &\propto \exp \left[ \frac{1}{2} \sum_{i,j} \op{c}_i^{\dagger} G_{i,j} \op{c}_j^{\dagger} \right] |0_1 \ldots 0_{4N} \rangle .
\end{align}

Having constructed $\mat{G}$, the brunt of the numerical effort now lies in representing $|I_N\rangle$.
This can be accomplished using matrix product states (MPS), which circumvents memory resource requirements growing exponentially with $N$.
Moreover, it has been recently demonstrated that for impurity models of the type considered here, the maximum entanglement entropy in $|I_N\rangle$ typically saturates to an area-law behavior as $N\to\infty$~\cite{Lerose2021, Thoenniss2022}.
This suggests that $|I_N\rangle$, and therefore the impurity dynamics, can be efficiently simulated using an MPS with low bond dimension, at least for some classes of impurity models.
The efficiency hinges crucially on the approach used to construct an MPS approximation of $|I_N\rangle$ (MPS-IF).
We now outline two methods for doing so: one which directly constructs each site tensor of the MPS-IF to produce an optimal low-rank MPS approximation, and another which reuses information from previous timesteps.

\subsection{Direct Construction}
A many-body MPS can be constructed from its site tensors by considering the overlaps of Schmidt states for two different bipartitions of $|I_N\rangle$~\cite{Schollwoeck2011}.
Here, one can leverage the fact that $|I_N\rangle$ is a Bardeen-Cooper-Schrieffer (BCS) state~\cite{Thoenniss2022} and therefore can be transformed into a Hartree-Fock state, for which an efficient method exists to construct its MPS representation directly~\cite{Petrica2021}.
The transformation into a Hartree-Fock state proceeds by finding the Bogoliubov quasiparticles, which can be done by diagonalizing the matrix $\mat{G}$ in \nnref{eq:IF-state-definition}{Eq.~(}{)} at a cost of $\bigO(N^3)$ for the $N$th timestep, and performing a particle-hole transformation. 
This allows for a direct construction of the Schmidt decomposition of any bipartition of $|I_N\rangle$ between sites $(\ell, \ell+1)$ as,
\begin{align}
\label{eq:schmidt-decomp-single-particle}
   |I_N\rangle &\propto \prod_{i=1}^{2N} \left[ \sqrt{\nu^{[\ell]}_{i}} \op{\phi}_{i, L}^{[\ell]\dagger} + \sqrt{1 - \nu^{[\ell]}_{i} } \op{\phi}_{i, R}^{[\ell]\dagger} \right] |0_{L}^{[\ell]}\rangle \otimes |0_{R}^{[\ell]}\rangle.
\end{align}
The $\{ \op{\phi}_{i,L/R}^{[\ell]\dagger} \}$ is a set of orthonormal single-particle operators acting on the left/right partition, and $\{ \nu^{[\ell]}_{i} \}$ are their associated eigenvalues~\cite{Peschel2012}.
Note that the left and right Schmidt states are guaranteed to be Hartree-Fock states.
We can directly pick out the $D$ most relevant Schmidt states without needing to directly construct them.
While we can do so at the single-particle level (i.e., approximating all but $\log_2 D$ values $\nu^{[\ell]}_{i}$ with 0 or 1), we will keep the $D$ Schmidt states of highest weight.

Having picked out the relevant Schmidt states, the site tensor $A^{[\ell] \sigma_{\ell}}$ at the $\ell$-th site is specified by the overlaps,
\begin{align}
    A^{[\ell] \sigma_{\ell}}_{\alpha_{\ell-1}, \alpha_{\ell}} = \Big( \langle \alpha_{\ell-1, L} | \otimes \langle \sigma_{\ell} | \Big) |\alpha_{\ell, L}\rangle,
\end{align}
where $\alpha_{\ell, L}$ labels the left Schmidt states of the bipartition at the $\ell$th bond.
Via Wick's theorem, the elements of the site tensor are straightforwardly found from determinants of overlaps of single particle states, e.g.\ $\langle 0_{L}^{[\ell]}| \op{\phi}_{i, L}^{[\ell-1]} \op{\phi}_{j, L}^{[\ell]\dagger} | 0_{L}^{[\ell]}\rangle$.
Note that the matrix of overlaps only needs to be computed once per site tensor, as the required determinants can be formed from its submatrices.
The cost of constructing all $\bigO(N)$ single particle operators $\{ \op{\phi}_{i,L/R}^{[\ell]\dagger} \}$ and the overlap matrix is $\bigO(N^3)$.
Furthermore, the evaluation of determinants for each site matrix can be sped up by the fact that certain orbitals are always occupied in all the considered Schmidt states.
The determinant calculation can be broken up into the product of the determinant of the occupied block with the determinant of its Schur complement.
In all, a single site tensor can be constructed at a cost of $\bigO(N^3 + D^2 N^2 \log_2 D)$~\cite{Petrica2021}, meaning that the cost of constructing the full MPS-IF at timestep $N$ is $\bigO(N^4 + D^2 N^3 \log_2 D)$. Note that the above procedure constructs each tensor independently, which leads to trivial parallelization, but the time may also be reduced if truncated tensors are constructed in the basis of preceding tensors.

\subsection{Iterative Construction}
In addition to using the generalized Gaussian nature of $|I_N\rangle$ to directly construct an MPS-IF, we can take an approach in the same spirit as similar methods for bosonic impurity problems, i.e.\ the time-evolving matrix product operator (TEMPO) reformulation of the QUAPI approach~\cite{Strathearn2018, Jorgensen2019, Gribben2022, Cygorek2022}.
In TEMPO, the MPS-IF can be propagated from the $(N-1)$th to $N$th timestep by contracting it with a layer of matrix product operators (MPOs), which contain information on how the impurity state at timestep $N$ correlates with the impurity state at all previous times.
To work in the same spirit, let us start from \nnref{eq:IF-state-definition}{Eq.~(}{)} and decompose the quadratic exponent into the form,
\begin{align}
  |I_N\rangle &= \prod_j \exp \left[ G_{j-1,j} \op{c}_{j-1}^{\dagger} \op{c}_j^{\dagger} \right] \cdots \exp \left[ G_{1,j} \op{c}_1^{\dagger} \op{c}_j^{\dagger} \right] |0\rangle \nonumber \\
  &= \prod_j \exp \left[ \Bigg( \sum_{i=1}^{j-1} \Delta G_{i,j} \op{c}_i^{\dagger} \Bigg) \op{c}_j^{\dagger} \right] |I_{N-1}\rangle,
\end{align}
where $\Delta \mat{G}$ denotes the change in $\mat{G}$ between timesteps $N-1$ and $N$~\footnote{In cases where all terms of the impurity-bath coupling can be simultaneously diagonalized, $\Delta \mat{G}$ is zero everywhere except for the rows/columns containing correlations between the ``newest'' impurity states $\op{c}_{4n-3}^{\dagger}, \ldots, \op{c}_{4n}^{\dagger}$ and the previous states.}.
Let us examine one grouping of operators at fixed $j$ and define for convenience $g_i \equiv \Delta G_{i,j} \equiv |g_i| e^{i \phi_i}$,
\begin{align}
\label{eq:iterative-layer}
\exp \left[ \Bigg( \sum_{i=1}^{j-1} g_i \op{c}_i^{\dagger} \Bigg) \op{c}_j^{\dagger} \right] &= 1 + \Bigg( \sum_{i=1}^{j-1} g_i \op{c}_i^{\dagger} \Bigg) \op{c}_j^{\dagger}.
\end{align}
This operator explicitly contains contains long-ranged couplings.
While there are more sophisticated ways to represent such an object as an MPO~\cite{Parker2020}, we will instead use a straightforward approach of decomposing it as a sequence of nearest-neighboring unitary transformations $\op{V}_{i-1,i}$.
These are defined such that they transform a linear combination of two fermions into a single effective fermion,
\begin{align*}
g_{i-1} \op{c}_{i-1}^{\dagger} + g_i \op{c}_{i}^{\dagger} = \op{V}_{i-1,i} \Big( e^{i \phi_{i-1}} \sqrt{|g_{i-1}|^2 + |g_{i}|^2} \,\, \op{c}_{i}^{\dagger} \Big) \op{V}_{i-1,i}^{\dagger},
\end{align*}
from which one finds
\begin{widetext}
\begin{gather}
  \op{V}_{i-1,i} = \exp \left[ i \left( \frac{\phi_{i} - \phi_{i-1}}{2} \right) (\op{c}_{i}^{\dagger} \op{c}_{i} - \hc) \right] \exp \left[ \left( \arctan \frac{|g_{i-1}|}{|g_{i}|} \right) (\op{c}_{i-1}^{\dagger} \op{c}_i - \hc) \right].
\end{gather}
\end{widetext}
Combining these transformations gives a representation of \nnref{eq:iterative-layer}{Eq.~(}{)} for a single $j$.
Constructing the MPS-IF at the $N$th timestep from the $(N-1)$th timestep requires the $\bigO(N)$ applications of \nnref{eq:iterative-layer}{Eq.~(}{)}, each of which involves $\bigO(N)$ 2-site gates.
To keep the bond dimension manageable, truncations via singular value decompositions are performed after each application of \nnref{eq:iterative-layer}{Eq.~(}{)} for each $j$.
In total, cost of the MPO-MPS applications and the SVD truncations makes this method's asymptotic cost $\bigO(D^3 N^2)$, for an iterative step (which must be performed $N$ times for the evolution).

It is evident from this construction that the computational requirements can be relaxed if one can neglect parts of $\mat{G}$ that do not change appreciably between consecutive timesteps, or if we impose a restriction on the memory length e.g.\ set $G_{i,j} = 0$ for $|i-j| > M$.
The latter case may be admissible given that $G_{i,j}$ is argued to decay algebraically with $|i-j|$, so long as the initial bath state is not critical~\cite{Thoenniss2022}.
Such a memory truncation would reduce the number of MPO-MPS applications from $(4N - 1)^2$ to $(4N)(2M-3) - (M^2 - M - 1)$.
Similarly, the number of SVDs is reduced from $(4N)(4N-1)/2$ to $(4N)(M-1) - \frac{M}{2}(M-1)$.
In all, the overall scaling for an iterative step would be reduced to $\bigO(D^3 M N)$.

\begin{figure}[hbp]
    \centering
    \includegraphics[width=\columnwidth]{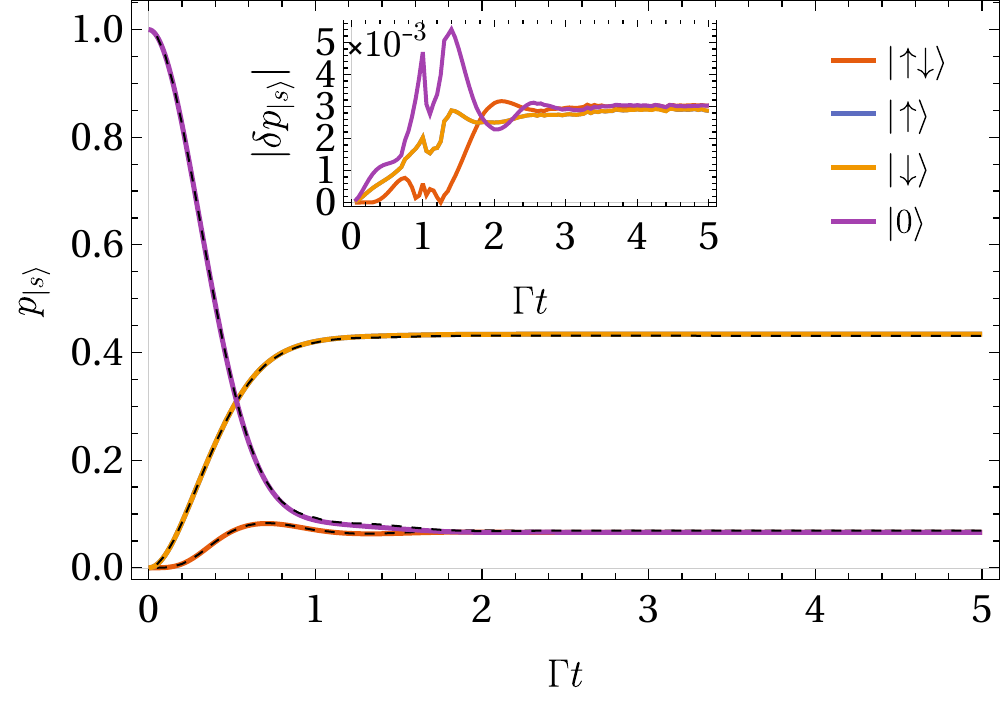}
    \caption{(Color online) Impurity populations $p_{|\psi\rangle} \equiv \langle \psi | \op{\rho} |\psi\rangle$ in the symmetric Anderson model with $U = 2.5\pi\Gamma$ and $\varepsilon_{\sigma} = -1.25\pi\Gamma$. The bath is initially at temperature $\Gamma \beta = 2$ and the impurity is initially unoccupied. Solid colored lines are results from~\cite{Kohn2022}, and black dashed lines are results with $D = 64$ using the direct construction of the MPS-IF with a timestep of $\Gamma \Delta t = 0.05$. \textbf{(inset)} Absolute deviations in the populations.}
    \label{fig:pops-64-santoro-comparison}
\end{figure}

\begin{figure*}
    \centering
    \includegraphics[width=\textwidth]{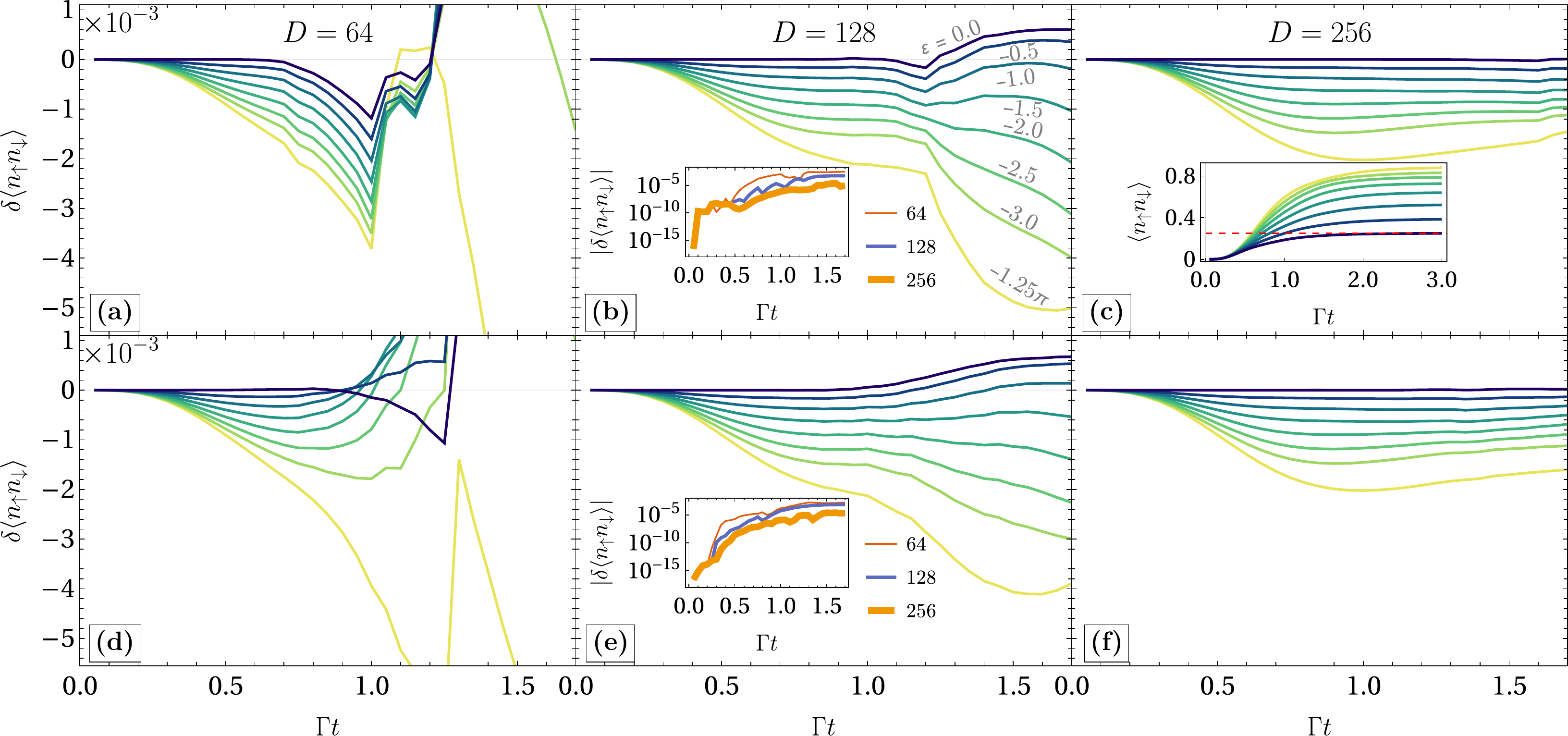}
    \caption{(Color online) Difference between the double occupancy computed from the MPS-IF with $\Gamma \Delta t = 0.05$, and its exact value from the solution of a differential equation~\cite{Note2}. The bath defined by \nnref{eq:semicircular-hybridization}{Eq.~}{} in the $U=0$ limit at temperature $\Gamma \beta = 2$, with an initially unoccupied impurity. The MPS-IF is constructed \textbf{(a-c)} directly and \textbf{(d-f)} iteratively, and $D = $ \textbf{(a,d)} $64$, \textbf{(b,e)} $128$, and \textbf{(c,f)} $256$. \textbf{(a and e, inset)} The absolute difference for $\varepsilon = 0$ across different bond dimensions. \textbf{(c, inset)} The exact double occupancies used as reference in the main panels. The red dashed line is the steady state value for $\varepsilon = 0$.}
    \label{fig:nonint-comparison}
\end{figure*}

\section{Results}

We consider the dynamics of the single impurity Anderson model for a hybridization function corresponding to the density of states of the $z \to \infty$ Bethe lattice,
\begin{align}
\label{eq:semicircular-hybridization}
    \Delta(\omega) = \Gamma \sqrt{W^2 - \omega^2}/\pi,
\end{align}
where $W = 10\Gamma$ and $\omega \in [-W, W]$.
The bath is initially equilibrated at a temperature $\Gamma \beta = 2$ and is decoupled from an unoccupied impurity $\op{\rho}(0) = |0\rangle\langle 0|$.
For $U = -2 \varepsilon_{\sigma} = 2.5\pi\Gamma$, we compare our results, generated using the direct and iterative constructions without memory length truncations, to those of the MPS time-dependent variational principle calculations of Kohn and Santoro~\cite{Kohn2022}.

In \nnref{fig:pops-64-santoro-comparison}{Fig.~}{} we show the quench dynamics of the populations, $p_{|\psi\rangle} \equiv \langle \psi | \op{\rho} | \psi \rangle$, from the directly constructed MPS-IF with a maximum bond dimension $D = 64$.
The timestep used is $\Gamma \Delta t = 0.05$.
As seen in the inset, the absolute difference between our results and those of Kohn and Santoro is on the order of $10^{-3}$ over the time range $\Gamma t \in [0, 5]$.
The deviations are generally larger than violations of the trace condition $\Tr \op{\rho}(t) = 1$~\footnote{Such violations are due solely to the approximation of $|I_N\rangle$ as a finite bond dimension MPS, since the IF represents unitary Trotterized dynamics.}.
Similar magnitudes of error are present from the iteratively constructed MPS-IF, which we show in~\cite{Note2}.
Notably, the dynamics from both construction methods do not perfectly coincide, due to the use of SVDs to compress the iteratively constructed MPS-IF.
For larger bond dimensions, this problem is exacerbated by the fact that the Schmidt values of $|I_N\rangle$ can fall below the precision of 64-bit floating point numbers so that the accuracy of the associated Schmidt vectors found by SVD cannot be guaranteed.
This, however, poses no issue for the direct construction method, for which the Schmidt values are found from products of the $\sqrt{\nu_{i}}$ and $\sqrt{1-\nu_{i}}$ in \nnref{eq:schmidt-decomp-single-particle}{Eq.~(}{)}.

The above considerations suggest that truncation error in the MPS construction must be carefully examined at the bond dimensions we use and must be considered in addition to Trotter error. 
We thus undertake a closer examination of these errors below.

\subsection{Convergence Analysis in the Non-Interacting Limit}
We begin our discussion by considering the dynamics for the case $U=0$, $\varepsilon_{\sigma} \neq 0$ since this admits exact numerical solutions~\footnote{Unlike for other methods, the dynamics of the non-interacting case are just as difficult to compute via the IF approach as they are for $U\neq 0$, since the presence or absence of $U$ does not alter the IF itself.}.
In \nnref{fig:nonint-comparison}{Fig.~}{} we show the deviation of the double occupancy $\langle \op{n}_{\uparrow} \op{n}_{\downarrow}\rangle$, as computed by the direct and iterative methods, from the exact values over a range of onsite energies $\varepsilon_{\sigma}$.
The most striking feature in these plots is a discontinuous rate of growth in the error $\delta\langle \op{n}_{\uparrow} \op{n}_{\downarrow}\rangle \equiv \langle \op{n}_{\uparrow} \op{n}_{\downarrow}\rangle - \langle \op{n}_{\uparrow} \op{n}_{\downarrow}\rangle_{\text{ref}}$, most prominently exhibited in simulations using MPS-IF with smaller bond dimensions and in the cases where the direct construction approach is used, \nnref{fig:nonint-comparison}{Fig.~}{(a,b,c)}.
Specifically, this behavior appears in \nnref{fig:nonint-comparison}{Fig.~}{b} at $\Gamma t^* \approx 1.2$.
In contrast, the nearly discontinuous behavior appears to be smoothed over in the iterative construction (\nnref{fig:nonint-comparison}{Fig.~}{(d,e,f)}), although the deviations in the double occupancy generally follow the same trends as in the directly constructed cases.
We observe $t^*$ to increase modestly with increasing bond dimension. For instance, at $D = 256$, $\Gamma t^* \approx 1.6$. Since the  deviations of the iteratively constructed method and direct method are qualitatively similar, below we focus on the behavior of the direct method.

\begin{figure}
    \centering
    \includegraphics[width=\columnwidth]{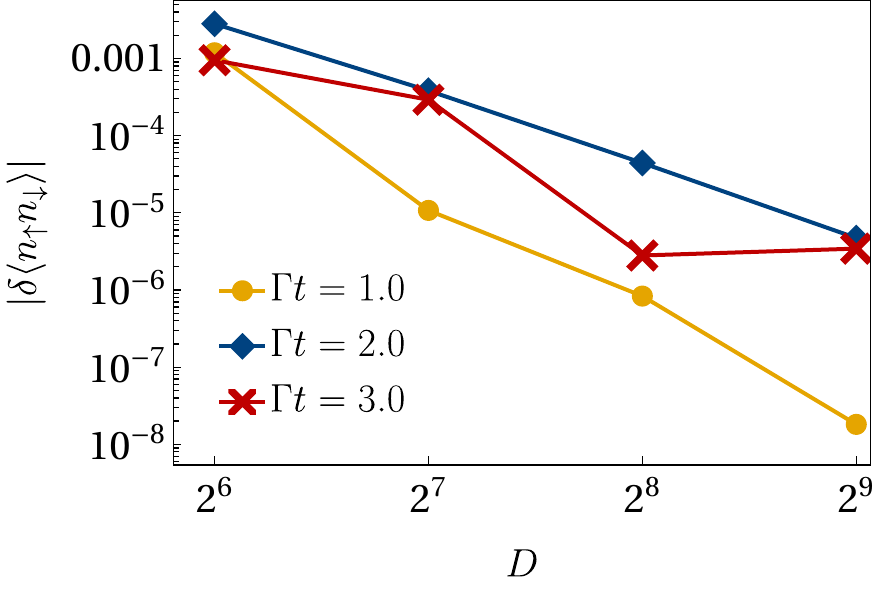}
    \caption{Absolute deviations from the exact double occupancy for $U = 0, \varepsilon = 0$ for various bond dimensions $D$ at fixed times approaching the steady state of the dynamics. The bath is initially at temperature $\Gamma \beta = 2$ and the impurity is initially unoccupied. The MPS-IF is constructed directly with a timestep of $\Gamma \Delta t = 0.05$.}
    \label{fig:nonint-comparison-longtime}
\end{figure}

\begin{figure*}
    \centering
    \includegraphics[width=\textwidth]{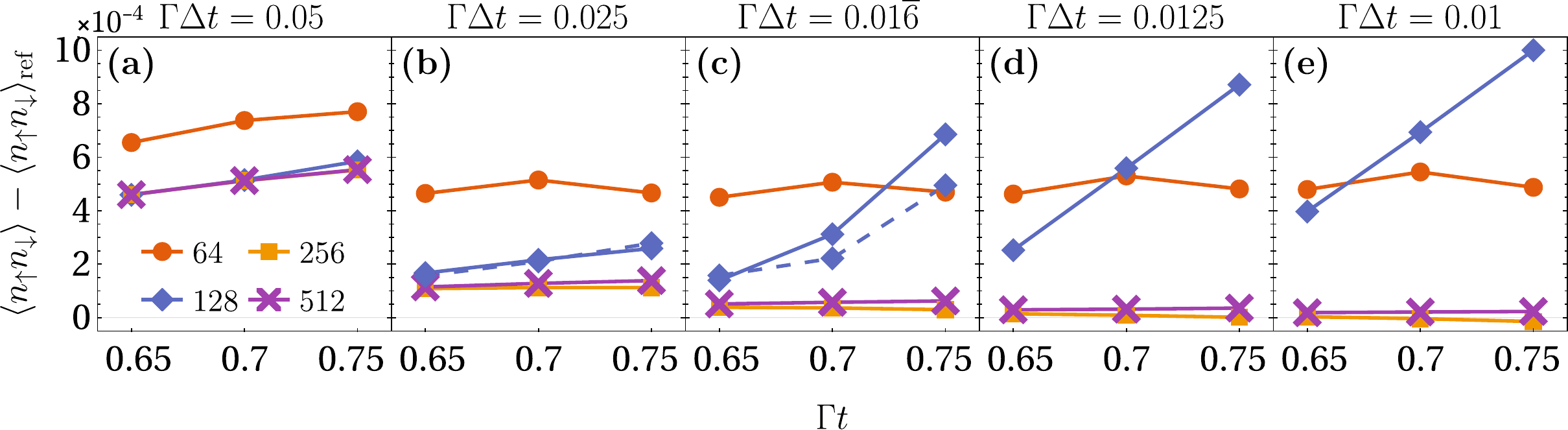}
    \caption{Convergence of the double occupancy with respect to $\Delta t$ and $D$, using the data from~\cite{Kohn2022} as the reference. The bath is initially at temperature $\Gamma\beta = 2$, as before. Data joined by solid (dashed) lines are obtained by the direct (iterative) construction method.}
    \label{fig:hump-comparison}
\end{figure*}

The error incurred by the MPS approximation can be isolated by examining the $\varepsilon = 0$ case, where there is no Trotter error.
These results are shown in the insets of \nnref{fig:nonint-comparison}{Fig.~}{(b,e)}.
It can be seen that the error increases dramatically with propagation time, growing by five decades over 20 time steps for $D = 128$ before saturating.
The error can be suppressed by increasing the bond dimension.
We generally see doing so decreases the error algebraically with $D$ (\nnref{fig:nonint-comparison-longtime}{Fig.~}{}).
Note that this holds at both intermediate ($\Gamma t \sim 1$) and long times ($\Gamma t \gtrsim 2$) for which the impurity dynamics is close to its steady state behavior (see bottom-most curve in the inset of \nnref{fig:nonint-comparison}{Fig.~}{c}).

At the same time, over the time range for which the dynamics are converged with respect to the available bond dimensions ($\Gamma t \lesssim 0.6$), the Trotter error 
scales as $(\Delta t)^{2}$~\cite{Note2}.
Thus, ensuring that the Trotter error is smaller than a tolerance $\delta$ implies that
$\Delta t \propto \delta^{1/2}$ and $N \propto \delta^{-1/2}$.
If the cost to reach a truncation error $\delta$ also scales asymptotically as a power law in $1/\delta$, then total cost to simulate the noninteracting impurity dynamics with the MPS-IF to the specified error tolerance will scale polynomially with $1/\delta$.

\subsection{Convergence Analysis in the Interacting Limit}

Given that the influence functional is independent of the details of the impurity Hamiltonian $\op{H}_1$, 
vestiges of the irregularities in the error from the non-interacting dynamics should also appear for the interacting dynamics.
This fact is useful, since it implies that by examining the non-interacting dynamics, we can anticipate points in time to focus convergence efforts for the interacting problem, e.g.\ close to the $t^{*}$ identified in the previous section.
However, the information gleaned from the non-interacting dynamics do not tell the full story.
For example, \nnref{fig:nonint-comparison}{Fig.~}{} shows that the double occupancy appears to be adequately converged with respect to $D$ around $\Gamma t \approx 0.7$.
Yet in the full problem with $U = -2 \varepsilon_{\sigma} = 2.5\pi \Gamma$ as shown in \nnref{fig:hump-comparison}{Fig.~}{}, we see that the error can behave rather differently as a function of both $D$ and $\Delta t$.
First, for $D \gtrsim 256$, we see that the deviations are most sensitive to the time step size, though there are small residual truncation errors independent of the Trotter error.
We surmise that the errors shown in \nnref{fig:hump-comparison}{Fig.~}{a} are mostly due to Trotter error, since $\delta\langle \op{n}_{\uparrow} \op{n}_{\downarrow} \rangle$ are identical for $D=256$ and $512$.
At the same time, the error for $D = 64$ is mostly unaffected by decreasing $\Delta t$, indicating that truncation errors are dominant.
We conclude that decreasing the time step size does not necessarily decrease the overall error given by the MPS-IF at fixed bond dimension.
As noted in Ref.~\cite{Thoenniss2022}, the half-cut von Neumann entanglement entropy of the MPS-IF vanishes as $\Delta t \to 0$.
However, this vanishing of the entanglement entropy, which is due to the scaling of the Schmidt values with $\Delta t$~\cite{SONNER2021168677, Note2}, does not imply that the accuracy of the MPS-IF approximation with fixed $D$ improves as $\Delta t \to 0$.
Our lack of rigorous understanding of the overall error is highlighted by the \textit{growth} of deviations with decreasing $\Delta t$ for $D = 128$, seen in \nnref{fig:hump-comparison}{Fig.~}{(c-e)}.
We observe a similar error behavior from the iterative construction, shown in the dashed lines in \nnref{fig:hump-comparison}{Fig.~}{c}.

\begin{figure}
    \centering
    \includegraphics[width=\columnwidth]{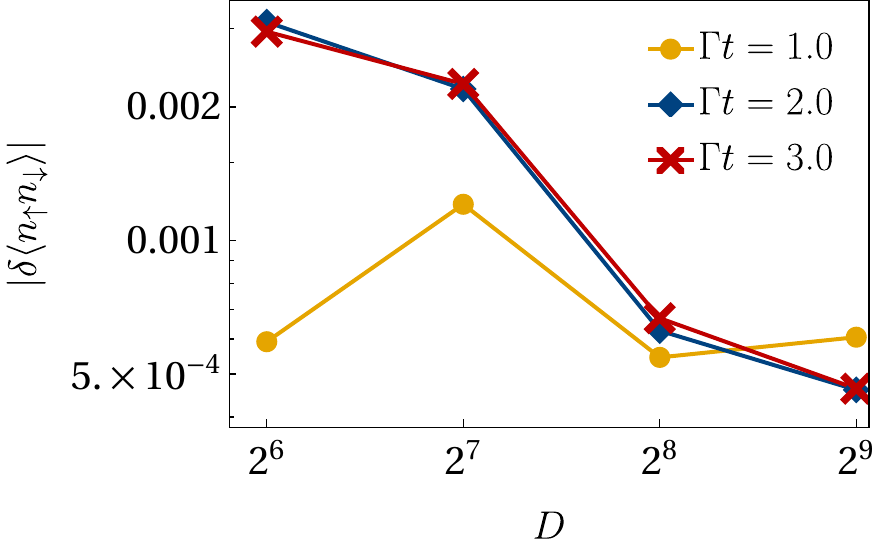}
    \caption{Comparison of the double occupancy in the $U \neq 0$ against results from Ref.~\cite{Kohn2022}, for various bond dimensions $D$.}
    \label{fig:full-comparison-longtime}
\end{figure}

Finally, we can make a similar comparison of the intermediate to long time dynamics as in \nnref{fig:nonint-comparison-longtime}{Fig.~}{} for the full $U \neq 0$ problem.
Restricting ourselves to the direct construction of the MPS-IF, we show in \nnref{fig:full-comparison-longtime}{Fig.~}{} the errors in the double occupancy across bond dimensions, holding $\Gamma \Delta t = 0.05$.
As in the non-interacting case, the errors at long times $\Gamma t \gtrsim 2$ suggest a power law decay with $D$ with a similar exponent.
In contrast, errors at the intermediate time $\Gamma t = 1$ do not exhibit such a decay, taking into account that the $D=128$ dynamics (\nnref{fig:hump-comparison}{Fig.~}{}) suffers from larger-than-expected errors.
Similar to \nnref{fig:hump-comparison}{Fig.~}{a}, it is likely that the errors at $\Gamma t = 1$ for $D \neq 128$ are dominated by Trotter errors.
Achieving convergence with respect to $D$ and $\Delta t$ may require first decreasing $\Delta t$ by a factor of $\sim 3$, cf.\ \nnref{fig:hump-comparison}{Fig.~}{(a, c)}.
While these results are suggestive of the long time ($\Gamma t \gtrsim 2$) dynamics being convergeable with polynomial effort for a specified error tolerance, 
larger bond dimensions will be required to reach this asymptotic convergence regime 
at intermediate times.

\section{Discussion}

In this paper we have presented two representations of fermionic bath influence functionals with matrix product states to simulate the real-time dynamics of the single-impurity Anderson model.
We have found that we can obtain good agreement with other real-time propagation methods with modest numerical effort.  
Importantly, we have the ability to treat arbitrary bath densities of state through the specification of its hybridization function~\cite{Note2}, and obtain systematically convergeable simulations of non-Markovian dynamics.
We stress here that unlike other dynamics methods, once we have constructed and saved the MPS-IF, calculations with different impurity Hamiltonians $\op{H}_1$ and different initial impurity states $\op{\rho}(0)$ can be performed trivially.  This allows for the treatment of, e.g., time-dependent forms of $\op{H}_1$ so that problems with external driving can be treated with no additional cost.
Furthermore, viewing the IF as a ``process tensor'' of an open quantum dynamics~\cite{Pollock2018} means that we can easily extract arbitrary multi-time impurity correlation functions within the same formalism.

We have also shown that the outstanding sources of error, Trotter error and truncation error of the MPS-IF, can likely be controlled with only polynomially growing resource requirements, but the two errors do not necessarily go hand in hand.
As presented, our approach should be readily generalizable to other impurity problems, and with some effort can be adapted as an impurity solver for DMFT.
These considerations, as well as modifications to improve computational efficiency, will be taken up in future work.

As it stands, we currently do not have a complete understanding of the major determinants of errors stemming from truncating the MPS-IF.
We anticipate that insights in this direction will help make the direct construction of the MPS-IF more efficient. Further explorations along these lines are forthcoming.

During preparation of this manuscript, we became aware of similar work by Thoenniss et al.~\cite{Thoenniss2022b}.

\begin{center}
    \textbf{Acknowledgements}
\end{center}
N.N.\ thanks Eran Rabani for assistance with calculations, which were performed using the ITensors library~\cite{ITensors}.
We thank Lucas Kohn and Giuseppe Santoro for providing data from their simulations.
This work was performed with support from the U.S.\ Department of Energy, Office of Science, Office of Advanced Scientific Computing Research, Scientific Discovery through Advanced Computing (SciDAC) program, under Award No. DE-SC0022088.
This research used resources of the National Energy Research Scientific Computing Center (NERSC), a U.S. Department of Energy Office of Science User Facility located at Lawrence Berkeley National Laboratory, operated under Contract No. DE-AC02-05CH11231.
The Flatiron Institute is a division of the Simons Foundation.
\bibliographystyle{apsrev4-2}
\bibliography{references}

\begin{thebibliography}{72}%
\makeatletter
\providecommand \@ifxundefined [1]{%
 \@ifx{#1\undefined}
}%
\providecommand \@ifnum [1]{%
 \ifnum #1\expandafter \@firstoftwo
 \else \expandafter \@secondoftwo
 \fi
}%
\providecommand \@ifx [1]{%
 \ifx #1\expandafter \@firstoftwo
 \else \expandafter \@secondoftwo
 \fi
}%
\providecommand \natexlab [1]{#1}%
\providecommand \enquote  [1]{``#1''}%
\providecommand \bibnamefont  [1]{#1}%
\providecommand \bibfnamefont [1]{#1}%
\providecommand \citenamefont [1]{#1}%
\providecommand \href@noop [0]{\@secondoftwo}%
\providecommand \href [0]{\begingroup \@sanitize@url \@href}%
\providecommand \@href[1]{\@@startlink{#1}\@@href}%
\providecommand \@@href[1]{\endgroup#1\@@endlink}%
\providecommand \@sanitize@url [0]{\catcode `\\12\catcode `\$12\catcode
  `\&12\catcode `\#12\catcode `\^12\catcode `\_12\catcode `\%12\relax}%
\providecommand \@@startlink[1]{}%
\providecommand \@@endlink[0]{}%
\providecommand \url  [0]{\begingroup\@sanitize@url \@url }%
\providecommand \@url [1]{\endgroup\@href {#1}{\urlprefix }}%
\providecommand \urlprefix  [0]{URL }%
\providecommand \Eprint [0]{\href }%
\providecommand \doibase [0]{https://doi.org/}%
\providecommand \selectlanguage [0]{\@gobble}%
\providecommand \bibinfo  [0]{\@secondoftwo}%
\providecommand \bibfield  [0]{\@secondoftwo}%
\providecommand \translation [1]{[#1]}%
\providecommand \BibitemOpen [0]{}%
\providecommand \bibitemStop [0]{}%
\providecommand \bibitemNoStop [0]{.\EOS\space}%
\providecommand \EOS [0]{\spacefactor3000\relax}%
\providecommand \BibitemShut  [1]{\csname bibitem#1\endcsname}%
\let\auto@bib@innerbib\@empty
\bibitem [{\citenamefont {Hewson}(1993)}]{Hewson1993}%
  \BibitemOpen
  \bibfield  {author} {\bibinfo {author} {\bibfnamefont {A.~C.}\ \bibnamefont
  {Hewson}},\ }\href {https://doi.org/10.1017/cbo9780511470752} {\emph
  {\bibinfo {title} {The Kondo Problem to Heavy Fermions}}}\ (\bibinfo
  {publisher} {Cambridge University Press},\ \bibinfo {year}
  {1993})\BibitemShut {NoStop}%
\bibitem [{\citenamefont {Leggett}\ \emph {et~al.}(1987)\citenamefont
  {Leggett}, \citenamefont {Chakravarty}, \citenamefont {Dorsey}, \citenamefont
  {Fisher}, \citenamefont {Garg},\ and\ \citenamefont {Zwerger}}]{Leggett1987}%
  \BibitemOpen
  \bibfield  {author} {\bibinfo {author} {\bibfnamefont {A.~J.}\ \bibnamefont
  {Leggett}}, \bibinfo {author} {\bibfnamefont {S.}~\bibnamefont
  {Chakravarty}}, \bibinfo {author} {\bibfnamefont {A.~T.}\ \bibnamefont
  {Dorsey}}, \bibinfo {author} {\bibfnamefont {M.~P.~A.}\ \bibnamefont
  {Fisher}}, \bibinfo {author} {\bibfnamefont {A.}~\bibnamefont {Garg}},\ and\
  \bibinfo {author} {\bibfnamefont {W.}~\bibnamefont {Zwerger}},\ }\href
  {https://doi.org/10.1103/RevModPhys.59.1} {\bibfield  {journal} {\bibinfo
  {journal} {Rev. Mod. Phys.}\ }\textbf {\bibinfo {volume} {59}},\ \bibinfo
  {pages} {1} (\bibinfo {year} {1987})}\BibitemShut {NoStop}%
\bibitem [{\citenamefont {Weiss}(2021)}]{Weiss2021}%
  \BibitemOpen
  \bibfield  {author} {\bibinfo {author} {\bibfnamefont {U.}~\bibnamefont
  {Weiss}},\ }\href {https://doi.org/10.1142/12402} {\emph {\bibinfo {title}
  {Quantum Dissipative Systems}}}\ (\bibinfo  {publisher} {World Scientific},\
  \bibinfo {year} {2021})\BibitemShut {NoStop}%
\bibitem [{\citenamefont {Georges}\ \emph {et~al.}(1996)\citenamefont
  {Georges}, \citenamefont {Kotliar}, \citenamefont {Krauth},\ and\
  \citenamefont {Rozenberg}}]{Georges1996}%
  \BibitemOpen
  \bibfield  {author} {\bibinfo {author} {\bibfnamefont {A.}~\bibnamefont
  {Georges}}, \bibinfo {author} {\bibfnamefont {G.}~\bibnamefont {Kotliar}},
  \bibinfo {author} {\bibfnamefont {W.}~\bibnamefont {Krauth}},\ and\ \bibinfo
  {author} {\bibfnamefont {M.~J.}\ \bibnamefont {Rozenberg}},\ }\href
  {https://doi.org/10.1103/RevModPhys.68.13} {\bibfield  {journal} {\bibinfo
  {journal} {Rev. Mod. Phys.}\ }\textbf {\bibinfo {volume} {68}},\ \bibinfo
  {pages} {13} (\bibinfo {year} {1996})}\BibitemShut {NoStop}%
\bibitem [{\citenamefont {Knizia}\ and\ \citenamefont
  {Chan}(2012)}]{Knizia2012}%
  \BibitemOpen
  \bibfield  {author} {\bibinfo {author} {\bibfnamefont {G.}~\bibnamefont
  {Knizia}}\ and\ \bibinfo {author} {\bibfnamefont {G.~K.-L.}\ \bibnamefont
  {Chan}},\ }\bibfield  {journal} {\bibinfo  {journal} {Physical Review
  Letters}\ }\textbf {\bibinfo {volume} {109}},\ \href
  {https://doi.org/10.1103/physrevlett.109.186404}
  {10.1103/physrevlett.109.186404} (\bibinfo {year} {2012})\BibitemShut
  {NoStop}%
\bibitem [{\citenamefont {Kretchmer}\ and\ \citenamefont
  {Chan}(2018)}]{Kretchmer2018}%
  \BibitemOpen
  \bibfield  {author} {\bibinfo {author} {\bibfnamefont {J.~S.}\ \bibnamefont
  {Kretchmer}}\ and\ \bibinfo {author} {\bibfnamefont {G.~K.-L.}\ \bibnamefont
  {Chan}},\ }\href {https://doi.org/10.1063/1.5012766} {\bibfield  {journal}
  {\bibinfo  {journal} {The Journal of Chemical Physics}\ }\textbf {\bibinfo
  {volume} {148}},\ \bibinfo {pages} {054108} (\bibinfo {year}
  {2018})}\BibitemShut {NoStop}%
\bibitem [{\citenamefont {Liebsch}\ and\ \citenamefont
  {Ishida}(2011)}]{Liebsch2011}%
  \BibitemOpen
  \bibfield  {author} {\bibinfo {author} {\bibfnamefont {A.}~\bibnamefont
  {Liebsch}}\ and\ \bibinfo {author} {\bibfnamefont {H.}~\bibnamefont
  {Ishida}},\ }\href {https://doi.org/10.1088/0953-8984/24/5/053201} {\bibfield
   {journal} {\bibinfo  {journal} {Journal of Physics: Condensed Matter}\
  }\textbf {\bibinfo {volume} {24}},\ \bibinfo {pages} {053201} (\bibinfo
  {year} {2011})}\BibitemShut {NoStop}%
\bibitem [{\citenamefont {Granath}\ and\ \citenamefont
  {Strand}(2012)}]{Granath2012}%
  \BibitemOpen
  \bibfield  {author} {\bibinfo {author} {\bibfnamefont {M.}~\bibnamefont
  {Granath}}\ and\ \bibinfo {author} {\bibfnamefont {H.~U.~R.}\ \bibnamefont
  {Strand}},\ }\href {https://doi.org/10.1103/PhysRevB.86.115111} {\bibfield
  {journal} {\bibinfo  {journal} {Phys. Rev. B}\ }\textbf {\bibinfo {volume}
  {86}},\ \bibinfo {pages} {115111} (\bibinfo {year} {2012})}\BibitemShut
  {NoStop}%
\bibitem [{\citenamefont {Lu}\ \emph {et~al.}(2014)\citenamefont {Lu},
  \citenamefont {H\"oppner}, \citenamefont {Gunnarsson},\ and\ \citenamefont
  {Haverkort}}]{Lu2014}%
  \BibitemOpen
  \bibfield  {author} {\bibinfo {author} {\bibfnamefont {Y.}~\bibnamefont
  {Lu}}, \bibinfo {author} {\bibfnamefont {M.}~\bibnamefont {H\"oppner}},
  \bibinfo {author} {\bibfnamefont {O.}~\bibnamefont {Gunnarsson}},\ and\
  \bibinfo {author} {\bibfnamefont {M.~W.}\ \bibnamefont {Haverkort}},\ }\href
  {https://doi.org/10.1103/PhysRevB.90.085102} {\bibfield  {journal} {\bibinfo
  {journal} {Phys. Rev. B}\ }\textbf {\bibinfo {volume} {90}},\ \bibinfo
  {pages} {085102} (\bibinfo {year} {2014})}\BibitemShut {NoStop}%
\bibitem [{\citenamefont {de~Vega}\ \emph {et~al.}(2015)\citenamefont
  {de~Vega}, \citenamefont {Schollw\"ock},\ and\ \citenamefont
  {Wolf}}]{deVega2015}%
  \BibitemOpen
  \bibfield  {author} {\bibinfo {author} {\bibfnamefont {I.}~\bibnamefont
  {de~Vega}}, \bibinfo {author} {\bibfnamefont {U.}~\bibnamefont
  {Schollw\"ock}},\ and\ \bibinfo {author} {\bibfnamefont {F.~A.}\ \bibnamefont
  {Wolf}},\ }\href {https://doi.org/10.1103/PhysRevB.92.155126} {\bibfield
  {journal} {\bibinfo  {journal} {Phys. Rev. B}\ }\textbf {\bibinfo {volume}
  {92}},\ \bibinfo {pages} {155126} (\bibinfo {year} {2015})}\BibitemShut
  {NoStop}%
\bibitem [{\citenamefont {G\"uttge}\ \emph {et~al.}(2013)\citenamefont
  {G\"uttge}, \citenamefont {Anders}, \citenamefont {Schollw\"ock},
  \citenamefont {Eidelstein},\ and\ \citenamefont {Schiller}}]{Guettge2013}%
  \BibitemOpen
  \bibfield  {author} {\bibinfo {author} {\bibfnamefont {F.}~\bibnamefont
  {G\"uttge}}, \bibinfo {author} {\bibfnamefont {F.~B.}\ \bibnamefont
  {Anders}}, \bibinfo {author} {\bibfnamefont {U.}~\bibnamefont
  {Schollw\"ock}}, \bibinfo {author} {\bibfnamefont {E.}~\bibnamefont
  {Eidelstein}},\ and\ \bibinfo {author} {\bibfnamefont {A.}~\bibnamefont
  {Schiller}},\ }\href {https://doi.org/10.1103/PhysRevB.87.115115} {\bibfield
  {journal} {\bibinfo  {journal} {Phys. Rev. B}\ }\textbf {\bibinfo {volume}
  {87}},\ \bibinfo {pages} {115115} (\bibinfo {year} {2013})}\BibitemShut
  {NoStop}%
\bibitem [{\citenamefont {Wolf}\ \emph {et~al.}(2014)\citenamefont {Wolf},
  \citenamefont {McCulloch}, \citenamefont {Parcollet},\ and\ \citenamefont
  {Schollw\"ock}}]{Wolf2014}%
  \BibitemOpen
  \bibfield  {author} {\bibinfo {author} {\bibfnamefont {F.~A.}\ \bibnamefont
  {Wolf}}, \bibinfo {author} {\bibfnamefont {I.~P.}\ \bibnamefont {McCulloch}},
  \bibinfo {author} {\bibfnamefont {O.}~\bibnamefont {Parcollet}},\ and\
  \bibinfo {author} {\bibfnamefont {U.}~\bibnamefont {Schollw\"ock}},\ }\href
  {https://doi.org/10.1103/PhysRevB.90.115124} {\bibfield  {journal} {\bibinfo
  {journal} {Phys. Rev. B}\ }\textbf {\bibinfo {volume} {90}},\ \bibinfo
  {pages} {115124} (\bibinfo {year} {2014})}\BibitemShut {NoStop}%
\bibitem [{\citenamefont {Ganahl}\ \emph {et~al.}(2015)\citenamefont {Ganahl},
  \citenamefont {Aichhorn}, \citenamefont {Evertz}, \citenamefont
  {Thunstr\"om}, \citenamefont {Held},\ and\ \citenamefont
  {Verstraete}}]{Ganahl2015}%
  \BibitemOpen
  \bibfield  {author} {\bibinfo {author} {\bibfnamefont {M.}~\bibnamefont
  {Ganahl}}, \bibinfo {author} {\bibfnamefont {M.}~\bibnamefont {Aichhorn}},
  \bibinfo {author} {\bibfnamefont {H.~G.}\ \bibnamefont {Evertz}}, \bibinfo
  {author} {\bibfnamefont {P.}~\bibnamefont {Thunstr\"om}}, \bibinfo {author}
  {\bibfnamefont {K.}~\bibnamefont {Held}},\ and\ \bibinfo {author}
  {\bibfnamefont {F.}~\bibnamefont {Verstraete}},\ }\href
  {https://doi.org/10.1103/PhysRevB.92.155132} {\bibfield  {journal} {\bibinfo
  {journal} {Phys. Rev. B}\ }\textbf {\bibinfo {volume} {92}},\ \bibinfo
  {pages} {155132} (\bibinfo {year} {2015})}\BibitemShut {NoStop}%
\bibitem [{\citenamefont {Bauernfeind}\ \emph {et~al.}(2017)\citenamefont
  {Bauernfeind}, \citenamefont {Zingl}, \citenamefont {Triebl}, \citenamefont
  {Aichhorn},\ and\ \citenamefont {Evertz}}]{Bauernfeind2017}%
  \BibitemOpen
  \bibfield  {author} {\bibinfo {author} {\bibfnamefont {D.}~\bibnamefont
  {Bauernfeind}}, \bibinfo {author} {\bibfnamefont {M.}~\bibnamefont {Zingl}},
  \bibinfo {author} {\bibfnamefont {R.}~\bibnamefont {Triebl}}, \bibinfo
  {author} {\bibfnamefont {M.}~\bibnamefont {Aichhorn}},\ and\ \bibinfo
  {author} {\bibfnamefont {H.~G.}\ \bibnamefont {Evertz}},\ }\href
  {https://doi.org/10.1103/PhysRevX.7.031013} {\bibfield  {journal} {\bibinfo
  {journal} {Phys. Rev. X}\ }\textbf {\bibinfo {volume} {7}},\ \bibinfo {pages}
  {031013} (\bibinfo {year} {2017})}\BibitemShut {NoStop}%
\bibitem [{\citenamefont {Bauernfeind}\ \emph {et~al.}(2019)\citenamefont
  {Bauernfeind}, \citenamefont {Aichhorn},\ and\ \citenamefont
  {Evertz}}]{Bauernfeind2019}%
  \BibitemOpen
  \bibfield  {author} {\bibinfo {author} {\bibfnamefont {D.}~\bibnamefont
  {Bauernfeind}}, \bibinfo {author} {\bibfnamefont {M.}~\bibnamefont
  {Aichhorn}},\ and\ \bibinfo {author} {\bibfnamefont {H.~G.}\ \bibnamefont
  {Evertz}},\ }\href@noop {} {} (\bibinfo {year} {2019}),\ \Eprint
  {https://arxiv.org/abs/arXiv:1906.09077} {arXiv:1906.09077} \BibitemShut
  {NoStop}%
\bibitem [{\citenamefont {Kohn}\ and\ \citenamefont
  {Santoro}(2022)}]{Kohn2022}%
  \BibitemOpen
  \bibfield  {author} {\bibinfo {author} {\bibfnamefont {L.}~\bibnamefont
  {Kohn}}\ and\ \bibinfo {author} {\bibfnamefont {G.~E.}\ \bibnamefont
  {Santoro}},\ }\href {https://doi.org/10.1088/1742-5468/ac729b} {\bibfield
  {journal} {\bibinfo  {journal} {Journal of Statistical Mechanics: Theory and
  Experiment}\ }\textbf {\bibinfo {volume} {2022}},\ \bibinfo {pages} {063102}
  (\bibinfo {year} {2022})}\BibitemShut {NoStop}%
\bibitem [{\citenamefont {M\"{u}hlbacher}\ and\ \citenamefont
  {Rabani}(2008)}]{Muehlbacher2008}%
  \BibitemOpen
  \bibfield  {author} {\bibinfo {author} {\bibfnamefont {L.}~\bibnamefont
  {M\"{u}hlbacher}}\ and\ \bibinfo {author} {\bibfnamefont {E.}~\bibnamefont
  {Rabani}},\ }\bibfield  {journal} {\bibinfo  {journal} {Physical Review
  Letters}\ }\textbf {\bibinfo {volume} {100}},\ \href
  {https://doi.org/10.1103/physrevlett.100.176403}
  {10.1103/physrevlett.100.176403} (\bibinfo {year} {2008})\BibitemShut
  {NoStop}%
\bibitem [{\citenamefont {Cohen}\ and\ \citenamefont
  {Rabani}(2011)}]{Cohen2011}%
  \BibitemOpen
  \bibfield  {author} {\bibinfo {author} {\bibfnamefont {G.}~\bibnamefont
  {Cohen}}\ and\ \bibinfo {author} {\bibfnamefont {E.}~\bibnamefont {Rabani}},\
  }\href {https://doi.org/10.1103/PhysRevB.84.075150} {\bibfield  {journal}
  {\bibinfo  {journal} {Phys. Rev. B}\ }\textbf {\bibinfo {volume} {84}},\
  \bibinfo {pages} {075150} (\bibinfo {year} {2011})}\BibitemShut {NoStop}%
\bibitem [{\citenamefont {Gull}\ \emph {et~al.}(2011)\citenamefont {Gull},
  \citenamefont {Millis}, \citenamefont {Lichtenstein}, \citenamefont
  {Rubtsov}, \citenamefont {Troyer},\ and\ \citenamefont {Werner}}]{Gull2011}%
  \BibitemOpen
  \bibfield  {author} {\bibinfo {author} {\bibfnamefont {E.}~\bibnamefont
  {Gull}}, \bibinfo {author} {\bibfnamefont {A.~J.}\ \bibnamefont {Millis}},
  \bibinfo {author} {\bibfnamefont {A.~I.}\ \bibnamefont {Lichtenstein}},
  \bibinfo {author} {\bibfnamefont {A.~N.}\ \bibnamefont {Rubtsov}}, \bibinfo
  {author} {\bibfnamefont {M.}~\bibnamefont {Troyer}},\ and\ \bibinfo {author}
  {\bibfnamefont {P.}~\bibnamefont {Werner}},\ }\href
  {https://doi.org/10.1103/revmodphys.83.349} {\bibfield  {journal} {\bibinfo
  {journal} {Reviews of Modern Physics}\ }\textbf {\bibinfo {volume} {83}},\
  \bibinfo {pages} {349} (\bibinfo {year} {2011})}\BibitemShut {NoStop}%
\bibitem [{\citenamefont {Gull}\ \emph {et~al.}(2010)\citenamefont {Gull},
  \citenamefont {Reichman},\ and\ \citenamefont {Millis}}]{Gull2010}%
  \BibitemOpen
  \bibfield  {author} {\bibinfo {author} {\bibfnamefont {E.}~\bibnamefont
  {Gull}}, \bibinfo {author} {\bibfnamefont {D.~R.}\ \bibnamefont {Reichman}},\
  and\ \bibinfo {author} {\bibfnamefont {A.~J.}\ \bibnamefont {Millis}},\
  }\bibfield  {journal} {\bibinfo  {journal} {Physical Review B}\ }\textbf
  {\bibinfo {volume} {82}},\ \href {https://doi.org/10.1103/physrevb.82.075109}
  {10.1103/physrevb.82.075109} (\bibinfo {year} {2010})\BibitemShut {NoStop}%
\bibitem [{\citenamefont {Cohen}\ \emph {et~al.}(2013)\citenamefont {Cohen},
  \citenamefont {Gull}, \citenamefont {Reichman}, \citenamefont {Millis},\ and\
  \citenamefont {Rabani}}]{Cohen2013}%
  \BibitemOpen
  \bibfield  {author} {\bibinfo {author} {\bibfnamefont {G.}~\bibnamefont
  {Cohen}}, \bibinfo {author} {\bibfnamefont {E.}~\bibnamefont {Gull}},
  \bibinfo {author} {\bibfnamefont {D.~R.}\ \bibnamefont {Reichman}}, \bibinfo
  {author} {\bibfnamefont {A.~J.}\ \bibnamefont {Millis}},\ and\ \bibinfo
  {author} {\bibfnamefont {E.}~\bibnamefont {Rabani}},\ }\href
  {https://doi.org/10.1103/PhysRevB.87.195108} {\bibfield  {journal} {\bibinfo
  {journal} {Phys. Rev. B}\ }\textbf {\bibinfo {volume} {87}},\ \bibinfo
  {pages} {195108} (\bibinfo {year} {2013})}\BibitemShut {NoStop}%
\bibitem [{\citenamefont {Cohen}\ \emph
  {et~al.}(2014{\natexlab{a}})\citenamefont {Cohen}, \citenamefont {Gull},
  \citenamefont {Reichman},\ and\ \citenamefont {Millis}}]{Cohen2014a}%
  \BibitemOpen
  \bibfield  {author} {\bibinfo {author} {\bibfnamefont {G.}~\bibnamefont
  {Cohen}}, \bibinfo {author} {\bibfnamefont {E.}~\bibnamefont {Gull}},
  \bibinfo {author} {\bibfnamefont {D.~R.}\ \bibnamefont {Reichman}},\ and\
  \bibinfo {author} {\bibfnamefont {A.~J.}\ \bibnamefont {Millis}},\ }\href
  {https://doi.org/10.1103/PhysRevLett.112.146802} {\bibfield  {journal}
  {\bibinfo  {journal} {Phys. Rev. Lett.}\ }\textbf {\bibinfo {volume} {112}},\
  \bibinfo {pages} {146802} (\bibinfo {year} {2014}{\natexlab{a}})}\BibitemShut
  {NoStop}%
\bibitem [{\citenamefont {Cohen}\ \emph
  {et~al.}(2014{\natexlab{b}})\citenamefont {Cohen}, \citenamefont {Reichman},
  \citenamefont {Millis},\ and\ \citenamefont {Gull}}]{Cohen2014b}%
  \BibitemOpen
  \bibfield  {author} {\bibinfo {author} {\bibfnamefont {G.}~\bibnamefont
  {Cohen}}, \bibinfo {author} {\bibfnamefont {D.~R.}\ \bibnamefont {Reichman}},
  \bibinfo {author} {\bibfnamefont {A.~J.}\ \bibnamefont {Millis}},\ and\
  \bibinfo {author} {\bibfnamefont {E.}~\bibnamefont {Gull}},\ }\bibfield
  {journal} {\bibinfo  {journal} {Physical Review B}\ }\textbf {\bibinfo
  {volume} {89}},\ \href {https://doi.org/10.1103/physrevb.89.115139}
  {10.1103/physrevb.89.115139} (\bibinfo {year}
  {2014}{\natexlab{b}})\BibitemShut {NoStop}%
\bibitem [{\citenamefont {Cohen}\ \emph {et~al.}(2015)\citenamefont {Cohen},
  \citenamefont {Gull}, \citenamefont {Reichman},\ and\ \citenamefont
  {Millis}}]{Cohen2015}%
  \BibitemOpen
  \bibfield  {author} {\bibinfo {author} {\bibfnamefont {G.}~\bibnamefont
  {Cohen}}, \bibinfo {author} {\bibfnamefont {E.}~\bibnamefont {Gull}},
  \bibinfo {author} {\bibfnamefont {D.~R.}\ \bibnamefont {Reichman}},\ and\
  \bibinfo {author} {\bibfnamefont {A.~J.}\ \bibnamefont {Millis}},\ }\href
  {https://doi.org/10.1103/PhysRevLett.115.266802} {\bibfield  {journal}
  {\bibinfo  {journal} {Phys. Rev. Lett.}\ }\textbf {\bibinfo {volume} {115}},\
  \bibinfo {pages} {266802} (\bibinfo {year} {2015})}\BibitemShut {NoStop}%
\bibitem [{\citenamefont {Boag}\ \emph {et~al.}(2018)\citenamefont {Boag},
  \citenamefont {Gull},\ and\ \citenamefont {Cohen}}]{Boag2018}%
  \BibitemOpen
  \bibfield  {author} {\bibinfo {author} {\bibfnamefont {A.}~\bibnamefont
  {Boag}}, \bibinfo {author} {\bibfnamefont {E.}~\bibnamefont {Gull}},\ and\
  \bibinfo {author} {\bibfnamefont {G.}~\bibnamefont {Cohen}},\ }\href
  {https://doi.org/10.1103/PhysRevB.98.115152} {\bibfield  {journal} {\bibinfo
  {journal} {Phys. Rev. B}\ }\textbf {\bibinfo {volume} {98}},\ \bibinfo
  {pages} {115152} (\bibinfo {year} {2018})}\BibitemShut {NoStop}%
\bibitem [{\citenamefont {Cai}\ \emph {et~al.}(2020)\citenamefont {Cai},
  \citenamefont {Lu},\ and\ \citenamefont {Yang}}]{Cai2020}%
  \BibitemOpen
  \bibfield  {author} {\bibinfo {author} {\bibfnamefont {Z.}~\bibnamefont
  {Cai}}, \bibinfo {author} {\bibfnamefont {J.}~\bibnamefont {Lu}},\ and\
  \bibinfo {author} {\bibfnamefont {S.}~\bibnamefont {Yang}},\ }\href@noop {}
  {} (\bibinfo {year} {2020}),\ \Eprint
  {https://arxiv.org/abs/arXiv:2006.07654} {arXiv:2006.07654} \BibitemShut
  {NoStop}%
\bibitem [{\citenamefont {Li}\ \emph {et~al.}(2022)\citenamefont {Li},
  \citenamefont {Yu}, \citenamefont {Gull},\ and\ \citenamefont
  {Cohen}}]{Li2022}%
  \BibitemOpen
  \bibfield  {author} {\bibinfo {author} {\bibfnamefont {J.}~\bibnamefont
  {Li}}, \bibinfo {author} {\bibfnamefont {Y.}~\bibnamefont {Yu}}, \bibinfo
  {author} {\bibfnamefont {E.}~\bibnamefont {Gull}},\ and\ \bibinfo {author}
  {\bibfnamefont {G.}~\bibnamefont {Cohen}},\ }\bibfield  {journal} {\bibinfo
  {journal} {Physical Review B}\ }\textbf {\bibinfo {volume} {105}},\ \href
  {https://doi.org/10.1103/physrevb.105.165133} {10.1103/physrevb.105.165133}
  (\bibinfo {year} {2022})\BibitemShut {NoStop}%
\bibitem [{\citenamefont {Makri}(1992)}]{Makri1992}%
  \BibitemOpen
  \bibfield  {author} {\bibinfo {author} {\bibfnamefont {N.}~\bibnamefont
  {Makri}},\ }\href {https://doi.org/10.1016/0009-2614(92)85654-s} {\bibfield
  {journal} {\bibinfo  {journal} {Chemical Physics Letters}\ }\textbf {\bibinfo
  {volume} {193}},\ \bibinfo {pages} {435} (\bibinfo {year}
  {1992})}\BibitemShut {NoStop}%
\bibitem [{\citenamefont {Makri}\ and\ \citenamefont
  {Makarov}(1995)}]{Makri1995a}%
  \BibitemOpen
  \bibfield  {author} {\bibinfo {author} {\bibfnamefont {N.}~\bibnamefont
  {Makri}}\ and\ \bibinfo {author} {\bibfnamefont {D.~E.}\ \bibnamefont
  {Makarov}},\ }\href {https://doi.org/10.1063/1.469508} {\bibfield  {journal}
  {\bibinfo  {journal} {The Journal of Chemical Physics}\ }\textbf {\bibinfo
  {volume} {102}},\ \bibinfo {pages} {4600} (\bibinfo {year}
  {1995})}\BibitemShut {NoStop}%
\bibitem [{\citenamefont {Mundinar}\ \emph {et~al.}(2022)\citenamefont
  {Mundinar}, \citenamefont {Hahn}, \citenamefont {K\"{o}nig},\ and\
  \citenamefont {Hucht}}]{Mundinar2022}%
  \BibitemOpen
  \bibfield  {author} {\bibinfo {author} {\bibfnamefont {S.}~\bibnamefont
  {Mundinar}}, \bibinfo {author} {\bibfnamefont {A.}~\bibnamefont {Hahn}},
  \bibinfo {author} {\bibfnamefont {J.}~\bibnamefont {K\"{o}nig}},\ and\
  \bibinfo {author} {\bibfnamefont {A.}~\bibnamefont {Hucht}},\ }\bibfield
  {journal} {\bibinfo  {journal} {Physical Review B}\ }\textbf {\bibinfo
  {volume} {106}},\ \href {https://doi.org/10.1103/physrevb.106.165427}
  {10.1103/physrevb.106.165427} (\bibinfo {year} {2022})\BibitemShut {NoStop}%
\bibitem [{\citenamefont {Tanimura}\ and\ \citenamefont
  {Kubo}(1989)}]{Tanimura1989}%
  \BibitemOpen
  \bibfield  {author} {\bibinfo {author} {\bibfnamefont {Y.}~\bibnamefont
  {Tanimura}}\ and\ \bibinfo {author} {\bibfnamefont {R.}~\bibnamefont
  {Kubo}},\ }\href {https://doi.org/10.1143/jpsj.58.101} {\bibfield  {journal}
  {\bibinfo  {journal} {Journal of the Physical Society of Japan}\ }\textbf
  {\bibinfo {volume} {58}},\ \bibinfo {pages} {101} (\bibinfo {year}
  {1989})}\BibitemShut {NoStop}%
\bibitem [{\citenamefont {Ishizaki}\ and\ \citenamefont
  {Tanimura}(2005)}]{Ishizaki2005}%
  \BibitemOpen
  \bibfield  {author} {\bibinfo {author} {\bibfnamefont {A.}~\bibnamefont
  {Ishizaki}}\ and\ \bibinfo {author} {\bibfnamefont {Y.}~\bibnamefont
  {Tanimura}},\ }\href {https://doi.org/10.1143/jpsj.74.3131} {\bibfield
  {journal} {\bibinfo  {journal} {Journal of the Physical Society of Japan}\
  }\textbf {\bibinfo {volume} {74}},\ \bibinfo {pages} {3131} (\bibinfo {year}
  {2005})}\BibitemShut {NoStop}%
\bibitem [{Note1()}]{Note1}%
  \BibitemOpen
  \bibinfo {note} {See, however~\cite {Haertle2013, Haertle2015, Jin2008,
  Zheng2009, Dong2014, Li2012, Cirio2022}}\BibitemShut {NoStop}%
\bibitem [{\citenamefont {Makri}(2017)}]{Makri2017}%
  \BibitemOpen
  \bibfield  {author} {\bibinfo {author} {\bibfnamefont {N.}~\bibnamefont
  {Makri}},\ }\href {https://doi.org/10.1063/1.4979197} {\bibfield  {journal}
  {\bibinfo  {journal} {The Journal of Chemical Physics}\ }\textbf {\bibinfo
  {volume} {146}},\ \bibinfo {pages} {134101} (\bibinfo {year}
  {2017})}\BibitemShut {NoStop}%
\bibitem [{\citenamefont {Makri}(2020)}]{Makri2020}%
  \BibitemOpen
  \bibfield  {author} {\bibinfo {author} {\bibfnamefont {N.}~\bibnamefont
  {Makri}},\ }\href {https://doi.org/10.1021/acs.jctc.0c00039} {\bibfield
  {journal} {\bibinfo  {journal} {Journal of Chemical Theory and Computation}\
  }\textbf {\bibinfo {volume} {16}},\ \bibinfo {pages} {4038} (\bibinfo {year}
  {2020})}\BibitemShut {NoStop}%
\bibitem [{\citenamefont {Weiss}\ \emph {et~al.}(2008)\citenamefont {Weiss},
  \citenamefont {Eckel}, \citenamefont {Thorwart},\ and\ \citenamefont
  {Egger}}]{Weiss2008}%
  \BibitemOpen
  \bibfield  {author} {\bibinfo {author} {\bibfnamefont {S.}~\bibnamefont
  {Weiss}}, \bibinfo {author} {\bibfnamefont {J.}~\bibnamefont {Eckel}},
  \bibinfo {author} {\bibfnamefont {M.}~\bibnamefont {Thorwart}},\ and\
  \bibinfo {author} {\bibfnamefont {R.}~\bibnamefont {Egger}},\ }\bibfield
  {journal} {\bibinfo  {journal} {Physical Review B}\ }\textbf {\bibinfo
  {volume} {77}},\ \href {https://doi.org/10.1103/physrevb.77.195316}
  {10.1103/physrevb.77.195316} (\bibinfo {year} {2008})\BibitemShut {NoStop}%
\bibitem [{\citenamefont {Segal}\ \emph {et~al.}(2010)\citenamefont {Segal},
  \citenamefont {Millis},\ and\ \citenamefont {Reichman}}]{Segal2010}%
  \BibitemOpen
  \bibfield  {author} {\bibinfo {author} {\bibfnamefont {D.}~\bibnamefont
  {Segal}}, \bibinfo {author} {\bibfnamefont {A.~J.}\ \bibnamefont {Millis}},\
  and\ \bibinfo {author} {\bibfnamefont {D.~R.}\ \bibnamefont {Reichman}},\
  }\bibfield  {journal} {\bibinfo  {journal} {Physical Review B}\ }\textbf
  {\bibinfo {volume} {82}},\ \href {https://doi.org/10.1103/physrevb.82.205323}
  {10.1103/physrevb.82.205323} (\bibinfo {year} {2010})\BibitemShut {NoStop}%
\bibitem [{\citenamefont {Ba\~nuls}\ \emph {et~al.}(2009)\citenamefont
  {Ba\~nuls}, \citenamefont {Hastings}, \citenamefont {Verstraete},\ and\
  \citenamefont {Cirac}}]{Banuls2009}%
  \BibitemOpen
  \bibfield  {author} {\bibinfo {author} {\bibfnamefont {M.~C.}\ \bibnamefont
  {Ba\~nuls}}, \bibinfo {author} {\bibfnamefont {M.~B.}\ \bibnamefont
  {Hastings}}, \bibinfo {author} {\bibfnamefont {F.}~\bibnamefont
  {Verstraete}},\ and\ \bibinfo {author} {\bibfnamefont {J.~I.}\ \bibnamefont
  {Cirac}},\ }\href {https://doi.org/10.1103/PhysRevLett.102.240603} {\bibfield
   {journal} {\bibinfo  {journal} {Phys. Rev. Lett.}\ }\textbf {\bibinfo
  {volume} {102}},\ \bibinfo {pages} {240603} (\bibinfo {year}
  {2009})}\BibitemShut {NoStop}%
\bibitem [{\citenamefont {Hastings}\ and\ \citenamefont
  {Mahajan}(2015)}]{Hastings2015}%
  \BibitemOpen
  \bibfield  {author} {\bibinfo {author} {\bibfnamefont {M.~B.}\ \bibnamefont
  {Hastings}}\ and\ \bibinfo {author} {\bibfnamefont {R.}~\bibnamefont
  {Mahajan}},\ }\href {https://doi.org/10.1103/PhysRevA.91.032306} {\bibfield
  {journal} {\bibinfo  {journal} {Phys. Rev. A}\ }\textbf {\bibinfo {volume}
  {91}},\ \bibinfo {pages} {032306} (\bibinfo {year} {2015})}\BibitemShut
  {NoStop}%
\bibitem [{\citenamefont {Tirrito}\ \emph {et~al.}(2022)\citenamefont
  {Tirrito}, \citenamefont {Robinson}, \citenamefont {Lewenstein},
  \citenamefont {Ran},\ and\ \citenamefont {Tagliacozzo}}]{Tirrito2018}%
  \BibitemOpen
  \bibfield  {author} {\bibinfo {author} {\bibfnamefont {E.}~\bibnamefont
  {Tirrito}}, \bibinfo {author} {\bibfnamefont {N.~J.}\ \bibnamefont
  {Robinson}}, \bibinfo {author} {\bibfnamefont {M.}~\bibnamefont
  {Lewenstein}}, \bibinfo {author} {\bibfnamefont {S.-J.}\ \bibnamefont
  {Ran}},\ and\ \bibinfo {author} {\bibfnamefont {L.}~\bibnamefont
  {Tagliacozzo}},\ }\href@noop {} {} (\bibinfo {year} {2022}),\ \Eprint
  {https://arxiv.org/abs/arXiv:1810.08050} {arXiv:1810.08050} \BibitemShut
  {NoStop}%
\bibitem [{\citenamefont {Ye}\ and\ \citenamefont {Chan}(2021)}]{Ye2021}%
  \BibitemOpen
  \bibfield  {author} {\bibinfo {author} {\bibfnamefont {E.}~\bibnamefont
  {Ye}}\ and\ \bibinfo {author} {\bibfnamefont {G.~K.-L.}\ \bibnamefont
  {Chan}},\ }\href {https://doi.org/10.1063/5.0047260} {\bibfield  {journal}
  {\bibinfo  {journal} {J. Chem. Phys.}\ }\textbf {\bibinfo {volume} {155}},\
  \bibinfo {pages} {044104} (\bibinfo {year} {2021})},\ \Eprint
  {https://arxiv.org/abs/https://doi.org/10.1063/5.0047260}
  {https://doi.org/10.1063/5.0047260} \BibitemShut {NoStop}%
\bibitem [{\citenamefont {Bose}\ and\ \citenamefont
  {Walters}(2021)}]{Bose2021}%
  \BibitemOpen
  \bibfield  {author} {\bibinfo {author} {\bibfnamefont {A.}~\bibnamefont
  {Bose}}\ and\ \bibinfo {author} {\bibfnamefont {P.~L.}\ \bibnamefont
  {Walters}},\ }\href@noop {} {} (\bibinfo {year} {2021}),\ \Eprint
  {https://arxiv.org/abs/arXiv:2106.12523} {arXiv:2106.12523} \BibitemShut
  {NoStop}%
\bibitem [{\citenamefont {Lerose}\ \emph {et~al.}(2021)\citenamefont {Lerose},
  \citenamefont {Sonner},\ and\ \citenamefont {Abanin}}]{Lerose2021}%
  \BibitemOpen
  \bibfield  {author} {\bibinfo {author} {\bibfnamefont {A.}~\bibnamefont
  {Lerose}}, \bibinfo {author} {\bibfnamefont {M.}~\bibnamefont {Sonner}},\
  and\ \bibinfo {author} {\bibfnamefont {D.~A.}\ \bibnamefont {Abanin}},\
  }\href {https://doi.org/10.1103/PhysRevB.104.035137} {\bibfield  {journal}
  {\bibinfo  {journal} {Phys. Rev. B}\ }\textbf {\bibinfo {volume} {104}},\
  \bibinfo {pages} {035137} (\bibinfo {year} {2021})}\BibitemShut {NoStop}%
\bibitem [{\citenamefont {Strathearn}\ \emph {et~al.}(2018)\citenamefont
  {Strathearn}, \citenamefont {Kirton}, \citenamefont {Kilda}, \citenamefont
  {Keeling},\ and\ \citenamefont {Lovett}}]{Strathearn2018}%
  \BibitemOpen
  \bibfield  {author} {\bibinfo {author} {\bibfnamefont {A.}~\bibnamefont
  {Strathearn}}, \bibinfo {author} {\bibfnamefont {P.}~\bibnamefont {Kirton}},
  \bibinfo {author} {\bibfnamefont {D.}~\bibnamefont {Kilda}}, \bibinfo
  {author} {\bibfnamefont {J.}~\bibnamefont {Keeling}},\ and\ \bibinfo {author}
  {\bibfnamefont {B.~W.}\ \bibnamefont {Lovett}},\ }\bibfield  {journal}
  {\bibinfo  {journal} {Nature Communications}\ }\textbf {\bibinfo {volume}
  {9}},\ \href {https://doi.org/10.1038/s41467-018-05617-3}
  {10.1038/s41467-018-05617-3} (\bibinfo {year} {2018})\BibitemShut {NoStop}%
\bibitem [{\citenamefont {J{\o}rgensen}\ and\ \citenamefont
  {Pollock}(2019)}]{Jorgensen2019}%
  \BibitemOpen
  \bibfield  {author} {\bibinfo {author} {\bibfnamefont {M.~R.}\ \bibnamefont
  {J{\o}rgensen}}\ and\ \bibinfo {author} {\bibfnamefont {F.~A.}\ \bibnamefont
  {Pollock}},\ }\bibfield  {journal} {\bibinfo  {journal} {Physical Review
  Letters}\ }\textbf {\bibinfo {volume} {123}},\ \href
  {https://doi.org/10.1103/physrevlett.123.240602}
  {10.1103/physrevlett.123.240602} (\bibinfo {year} {2019})\BibitemShut
  {NoStop}%
\bibitem [{\citenamefont {Bose}\ and\ \citenamefont
  {Walters}(2022)}]{Bose2022a}%
  \BibitemOpen
  \bibfield  {author} {\bibinfo {author} {\bibfnamefont {A.}~\bibnamefont
  {Bose}}\ and\ \bibinfo {author} {\bibfnamefont {P.~L.}\ \bibnamefont
  {Walters}},\ }\href {https://doi.org/10.1063/5.0073234} {\bibfield  {journal}
  {\bibinfo  {journal} {The Journal of Chemical Physics}\ }\textbf {\bibinfo
  {volume} {156}},\ \bibinfo {pages} {024101} (\bibinfo {year}
  {2022})}\BibitemShut {NoStop}%
\bibitem [{\citenamefont {Bose}(2022)}]{Bose2022b}%
  \BibitemOpen
  \bibfield  {author} {\bibinfo {author} {\bibfnamefont {A.}~\bibnamefont
  {Bose}},\ }\href {https://doi.org/10.1103/PhysRevB.105.024309} {\bibfield
  {journal} {\bibinfo  {journal} {Phys. Rev. B}\ }\textbf {\bibinfo {volume}
  {105}},\ \bibinfo {pages} {024309} (\bibinfo {year} {2022})}\BibitemShut
  {NoStop}%
\bibitem [{\citenamefont {Gribben}\ \emph {et~al.}(2022)\citenamefont
  {Gribben}, \citenamefont {Rouse}, \citenamefont {Iles-Smith}, \citenamefont
  {Strathearn}, \citenamefont {Maguire}, \citenamefont {Kirton}, \citenamefont
  {Nazir}, \citenamefont {Gauger},\ and\ \citenamefont {Lovett}}]{Gribben2022}%
  \BibitemOpen
  \bibfield  {author} {\bibinfo {author} {\bibfnamefont {D.}~\bibnamefont
  {Gribben}}, \bibinfo {author} {\bibfnamefont {D.~M.}\ \bibnamefont {Rouse}},
  \bibinfo {author} {\bibfnamefont {J.}~\bibnamefont {Iles-Smith}}, \bibinfo
  {author} {\bibfnamefont {A.}~\bibnamefont {Strathearn}}, \bibinfo {author}
  {\bibfnamefont {H.}~\bibnamefont {Maguire}}, \bibinfo {author} {\bibfnamefont
  {P.}~\bibnamefont {Kirton}}, \bibinfo {author} {\bibfnamefont
  {A.}~\bibnamefont {Nazir}}, \bibinfo {author} {\bibfnamefont {E.~M.}\
  \bibnamefont {Gauger}},\ and\ \bibinfo {author} {\bibfnamefont {B.~W.}\
  \bibnamefont {Lovett}},\ }\href {https://doi.org/10.1103/PRXQuantum.3.010321}
  {\bibfield  {journal} {\bibinfo  {journal} {PRX Quantum}\ }\textbf {\bibinfo
  {volume} {3}},\ \bibinfo {pages} {010321} (\bibinfo {year}
  {2022})}\BibitemShut {NoStop}%
\bibitem [{\citenamefont {Cygorek}\ \emph {et~al.}(2022)\citenamefont
  {Cygorek}, \citenamefont {Cosacchi}, \citenamefont {Vagov}, \citenamefont
  {Axt}, \citenamefont {Lovett}, \citenamefont {Keeling},\ and\ \citenamefont
  {Gauger}}]{Cygorek2022}%
  \BibitemOpen
  \bibfield  {author} {\bibinfo {author} {\bibfnamefont {M.}~\bibnamefont
  {Cygorek}}, \bibinfo {author} {\bibfnamefont {M.}~\bibnamefont {Cosacchi}},
  \bibinfo {author} {\bibfnamefont {A.}~\bibnamefont {Vagov}}, \bibinfo
  {author} {\bibfnamefont {V.~M.}\ \bibnamefont {Axt}}, \bibinfo {author}
  {\bibfnamefont {B.~W.}\ \bibnamefont {Lovett}}, \bibinfo {author}
  {\bibfnamefont {J.}~\bibnamefont {Keeling}},\ and\ \bibinfo {author}
  {\bibfnamefont {E.~M.}\ \bibnamefont {Gauger}},\ }\href
  {https://doi.org/10.1038/s41567-022-01544-9} {\bibfield  {journal} {\bibinfo
  {journal} {Nature Physics}\ }\textbf {\bibinfo {volume} {18}},\ \bibinfo
  {pages} {662} (\bibinfo {year} {2022})}\BibitemShut {NoStop}%
\bibitem [{\citenamefont {Thoenniss}\ \emph
  {et~al.}(2022{\natexlab{a}})\citenamefont {Thoenniss}, \citenamefont
  {Lerose},\ and\ \citenamefont {Abanin}}]{Thoenniss2022}%
  \BibitemOpen
  \bibfield  {author} {\bibinfo {author} {\bibfnamefont {J.}~\bibnamefont
  {Thoenniss}}, \bibinfo {author} {\bibfnamefont {A.}~\bibnamefont {Lerose}},\
  and\ \bibinfo {author} {\bibfnamefont {D.~A.}\ \bibnamefont {Abanin}},\
  }\href@noop {} {} (\bibinfo {year} {2022}{\natexlab{a}}),\ \Eprint
  {https://arxiv.org/abs/arXiv:2205.04995} {arXiv:2205.04995} \BibitemShut
  {NoStop}%
\bibitem [{\citenamefont {Sonner}\ \emph {et~al.}(2021)\citenamefont {Sonner},
  \citenamefont {Lerose},\ and\ \citenamefont {Abanin}}]{SONNER2021168677}%
  \BibitemOpen
  \bibfield  {author} {\bibinfo {author} {\bibfnamefont {M.}~\bibnamefont
  {Sonner}}, \bibinfo {author} {\bibfnamefont {A.}~\bibnamefont {Lerose}},\
  and\ \bibinfo {author} {\bibfnamefont {D.~A.}\ \bibnamefont {Abanin}},\
  }\href {https://doi.org/https://doi.org/10.1016/j.aop.2021.168677} {\bibfield
   {journal} {\bibinfo  {journal} {Annals of Physics}\ }\textbf {\bibinfo
  {volume} {435}},\ \bibinfo {pages} {168677} (\bibinfo {year} {2021})},\
  \bibinfo {note} {special issue on Philip W. Anderson}\BibitemShut {NoStop}%
\bibitem [{Note2()}]{Note2}%
  \BibitemOpen
  \bibinfo {note} {We note that the types of correlated initial conditions
  relevant for the calculation of equilibrium correlation (Green's) functions
  can be straightforwardly generated from imaginary time evolutions within the
  same formalism~\cite {Shao2002}.}\BibitemShut {Stop}%
\bibitem [{Note3()}]{Note3}%
  \BibitemOpen
  \bibinfo {note} {See Supplementary Materials at [] for additional data, along
  with details of the derivation of the IF and the computation using the
  MPS-IF.}\BibitemShut {Stop}%
\bibitem [{\citenamefont {Gu}\ \emph {et~al.}(2010)\citenamefont {Gu},
  \citenamefont {Verstraete},\ and\ \citenamefont {Wen}}]{Gu2010}%
  \BibitemOpen
  \bibfield  {author} {\bibinfo {author} {\bibfnamefont {Z.-C.}\ \bibnamefont
  {Gu}}, \bibinfo {author} {\bibfnamefont {F.}~\bibnamefont {Verstraete}},\
  and\ \bibinfo {author} {\bibfnamefont {X.-G.}\ \bibnamefont {Wen}},\
  }\href@noop {} {} (\bibinfo {year} {2010}),\ \Eprint
  {https://arxiv.org/abs/arXiv:1004.2563} {arXiv:1004.2563} \BibitemShut
  {NoStop}%
\bibitem [{\citenamefont {Schollw\"{o}ck}(2011)}]{Schollwoeck2011}%
  \BibitemOpen
  \bibfield  {author} {\bibinfo {author} {\bibfnamefont {U.}~\bibnamefont
  {Schollw\"{o}ck}},\ }\href {https://doi.org/10.1016/j.aop.2010.09.012}
  {\bibfield  {journal} {\bibinfo  {journal} {Annals of Physics}\ }\textbf
  {\bibinfo {volume} {326}},\ \bibinfo {pages} {96} (\bibinfo {year}
  {2011})}\BibitemShut {NoStop}%
\bibitem [{\citenamefont {Petrica}\ \emph {et~al.}(2021)\citenamefont
  {Petrica}, \citenamefont {Zheng}, \citenamefont {Chan},\ and\ \citenamefont
  {Clark}}]{Petrica2021}%
  \BibitemOpen
  \bibfield  {author} {\bibinfo {author} {\bibfnamefont {G.}~\bibnamefont
  {Petrica}}, \bibinfo {author} {\bibfnamefont {B.-X.}\ \bibnamefont {Zheng}},
  \bibinfo {author} {\bibfnamefont {G.~K.-L.}\ \bibnamefont {Chan}},\ and\
  \bibinfo {author} {\bibfnamefont {B.~K.}\ \bibnamefont {Clark}},\ }\href
  {https://doi.org/10.1103/PhysRevB.103.125161} {\bibfield  {journal} {\bibinfo
   {journal} {Phys. Rev. B}\ }\textbf {\bibinfo {volume} {103}},\ \bibinfo
  {pages} {125161} (\bibinfo {year} {2021})}\BibitemShut {NoStop}%
\bibitem [{\citenamefont {Peschel}(2012)}]{Peschel2012}%
  \BibitemOpen
  \bibfield  {author} {\bibinfo {author} {\bibfnamefont {I.}~\bibnamefont
  {Peschel}},\ }\href {https://doi.org/10.1007/s13538-012-0074-1} {\bibfield
  {journal} {\bibinfo  {journal} {Brazilian Journal of Physics}\ }\textbf
  {\bibinfo {volume} {42}},\ \bibinfo {pages} {267} (\bibinfo {year}
  {2012})}\BibitemShut {NoStop}%
\bibitem [{Note4()}]{Note4}%
  \BibitemOpen
  \bibinfo {note} {In cases where all terms of the impurity-bath coupling can
  be simultaneously diagonalized, $\Delta \protect \pmb {G}$ is zero everywhere
  except for the rows/columns containing correlations between the ``newest''
  impurity states $\protect \hat {c}_{4n-3}^{\dagger }, \protect \ldots ,
  \protect \hat {c}_{4n}^{\dagger }$ and the previous states.}\BibitemShut
  {Stop}%
\bibitem [{\citenamefont {Parker}\ \emph {et~al.}(2020)\citenamefont {Parker},
  \citenamefont {Cao},\ and\ \citenamefont {Zaletel}}]{Parker2020}%
  \BibitemOpen
  \bibfield  {author} {\bibinfo {author} {\bibfnamefont {D.~E.}\ \bibnamefont
  {Parker}}, \bibinfo {author} {\bibfnamefont {X.}~\bibnamefont {Cao}},\ and\
  \bibinfo {author} {\bibfnamefont {M.~P.}\ \bibnamefont {Zaletel}},\ }\href
  {https://doi.org/10.1103/PhysRevB.102.035147} {\bibfield  {journal} {\bibinfo
   {journal} {Phys. Rev. B}\ }\textbf {\bibinfo {volume} {102}},\ \bibinfo
  {pages} {035147} (\bibinfo {year} {2020})}\BibitemShut {NoStop}%
\bibitem [{Note5()}]{Note5}%
  \BibitemOpen
  \bibinfo {note} {Such violations are due solely to the approximation of
  $|I_N\rangle $ as a finite bond dimension MPS, since the IF represents
  unitary Trotterized dynamics.}\BibitemShut {Stop}%
\bibitem [{Note6()}]{Note6}%
  \BibitemOpen
  \bibinfo {note} {Unlike for other methods, the dynamics of the
  non-interacting case are just as difficult to compute via the IF approach as
  they are for $U\protect \neq 0$, since the presence or absence of $U$ does
  not alter the IF itself.}\BibitemShut {Stop}%
\bibitem [{\citenamefont {Pollock}\ \emph {et~al.}(2018)\citenamefont
  {Pollock}, \citenamefont {Rodr\'{\i}guez-Rosario}, \citenamefont
  {Frauenheim}, \citenamefont {Paternostro},\ and\ \citenamefont
  {Modi}}]{Pollock2018}%
  \BibitemOpen
  \bibfield  {author} {\bibinfo {author} {\bibfnamefont {F.~A.}\ \bibnamefont
  {Pollock}}, \bibinfo {author} {\bibfnamefont {C.}~\bibnamefont
  {Rodr\'{\i}guez-Rosario}}, \bibinfo {author} {\bibfnamefont {T.}~\bibnamefont
  {Frauenheim}}, \bibinfo {author} {\bibfnamefont {M.}~\bibnamefont
  {Paternostro}},\ and\ \bibinfo {author} {\bibfnamefont {K.}~\bibnamefont
  {Modi}},\ }\href {https://doi.org/10.1103/PhysRevA.97.012127} {\bibfield
  {journal} {\bibinfo  {journal} {Phys. Rev. A}\ }\textbf {\bibinfo {volume}
  {97}},\ \bibinfo {pages} {012127} (\bibinfo {year} {2018})}\BibitemShut
  {NoStop}%
\bibitem [{\citenamefont {Thoenniss}\ \emph
  {et~al.}(2022{\natexlab{b}})\citenamefont {Thoenniss}, \citenamefont
  {Sonner}, \citenamefont {Lerose},\ and\ \citenamefont
  {Abanin}}]{Thoenniss2022b}%
  \BibitemOpen
  \bibfield  {author} {\bibinfo {author} {\bibfnamefont {J.}~\bibnamefont
  {Thoenniss}}, \bibinfo {author} {\bibfnamefont {M.}~\bibnamefont {Sonner}},
  \bibinfo {author} {\bibfnamefont {A.}~\bibnamefont {Lerose}},\ and\ \bibinfo
  {author} {\bibfnamefont {D.~A.}\ \bibnamefont {Abanin}},\ }\href@noop {}
  {\bibinfo {title} {An efficient method for quantum impurity problems out of
  equilibrium}} (\bibinfo {year} {2022}{\natexlab{b}}),\ \Eprint
  {https://arxiv.org/abs/arXiv:2211.10272} {arXiv:2211.10272} \BibitemShut
  {NoStop}%
\bibitem [{\citenamefont {Fishman}\ \emph {et~al.}(2022)\citenamefont
  {Fishman}, \citenamefont {White},\ and\ \citenamefont
  {Stoudenmire}}]{ITensors}%
  \BibitemOpen
  \bibfield  {author} {\bibinfo {author} {\bibfnamefont {M.}~\bibnamefont
  {Fishman}}, \bibinfo {author} {\bibfnamefont {S.~R.}\ \bibnamefont {White}},\
  and\ \bibinfo {author} {\bibfnamefont {E.~M.}\ \bibnamefont {Stoudenmire}},\
  }\href {https://doi.org/10.21468/SciPostPhysCodeb.4} {\bibfield  {journal}
  {\bibinfo  {journal} {SciPost Phys. Codebases}\ ,\ \bibinfo {pages} {4}}
  (\bibinfo {year} {2022})}\BibitemShut {NoStop}%
\bibitem [{\citenamefont {H\"artle}\ \emph {et~al.}(2013)\citenamefont
  {H\"artle}, \citenamefont {Cohen}, \citenamefont {Reichman},\ and\
  \citenamefont {Millis}}]{Haertle2013}%
  \BibitemOpen
  \bibfield  {author} {\bibinfo {author} {\bibfnamefont {R.}~\bibnamefont
  {H\"artle}}, \bibinfo {author} {\bibfnamefont {G.}~\bibnamefont {Cohen}},
  \bibinfo {author} {\bibfnamefont {D.~R.}\ \bibnamefont {Reichman}},\ and\
  \bibinfo {author} {\bibfnamefont {A.~J.}\ \bibnamefont {Millis}},\ }\href
  {https://doi.org/10.1103/PhysRevB.88.235426} {\bibfield  {journal} {\bibinfo
  {journal} {Phys. Rev. B}\ }\textbf {\bibinfo {volume} {88}},\ \bibinfo
  {pages} {235426} (\bibinfo {year} {2013})}\BibitemShut {NoStop}%
\bibitem [{\citenamefont {H\"artle}\ \emph {et~al.}(2015)\citenamefont
  {H\"artle}, \citenamefont {Cohen}, \citenamefont {Reichman},\ and\
  \citenamefont {Millis}}]{Haertle2015}%
  \BibitemOpen
  \bibfield  {author} {\bibinfo {author} {\bibfnamefont {R.}~\bibnamefont
  {H\"artle}}, \bibinfo {author} {\bibfnamefont {G.}~\bibnamefont {Cohen}},
  \bibinfo {author} {\bibfnamefont {D.~R.}\ \bibnamefont {Reichman}},\ and\
  \bibinfo {author} {\bibfnamefont {A.~J.}\ \bibnamefont {Millis}},\ }\href
  {https://doi.org/10.1103/PhysRevB.92.085430} {\bibfield  {journal} {\bibinfo
  {journal} {Phys. Rev. B}\ }\textbf {\bibinfo {volume} {92}},\ \bibinfo
  {pages} {085430} (\bibinfo {year} {2015})}\BibitemShut {NoStop}%
\bibitem [{\citenamefont {Jin}\ \emph {et~al.}(2008)\citenamefont {Jin},
  \citenamefont {Zheng},\ and\ \citenamefont {Yan}}]{Jin2008}%
  \BibitemOpen
  \bibfield  {author} {\bibinfo {author} {\bibfnamefont {J.}~\bibnamefont
  {Jin}}, \bibinfo {author} {\bibfnamefont {X.}~\bibnamefont {Zheng}},\ and\
  \bibinfo {author} {\bibfnamefont {Y.}~\bibnamefont {Yan}},\ }\href
  {https://doi.org/10.1063/1.2938087} {\bibfield  {journal} {\bibinfo
  {journal} {The Journal of Chemical Physics}\ }\textbf {\bibinfo {volume}
  {128}},\ \bibinfo {pages} {234703} (\bibinfo {year} {2008})}\BibitemShut
  {NoStop}%
\bibitem [{\citenamefont {Zheng}\ \emph {et~al.}(2009)\citenamefont {Zheng},
  \citenamefont {Jin}, \citenamefont {Welack}, \citenamefont {Luo},\ and\
  \citenamefont {Yan}}]{Zheng2009}%
  \BibitemOpen
  \bibfield  {author} {\bibinfo {author} {\bibfnamefont {X.}~\bibnamefont
  {Zheng}}, \bibinfo {author} {\bibfnamefont {J.}~\bibnamefont {Jin}}, \bibinfo
  {author} {\bibfnamefont {S.}~\bibnamefont {Welack}}, \bibinfo {author}
  {\bibfnamefont {M.}~\bibnamefont {Luo}},\ and\ \bibinfo {author}
  {\bibfnamefont {Y.}~\bibnamefont {Yan}},\ }\href
  {https://doi.org/10.1063/1.3123526} {\bibfield  {journal} {\bibinfo
  {journal} {The Journal of Chemical Physics}\ }\textbf {\bibinfo {volume}
  {130}},\ \bibinfo {pages} {164708} (\bibinfo {year} {2009})}\BibitemShut
  {NoStop}%
\bibitem [{\citenamefont {Hou}\ \emph {et~al.}(2014)\citenamefont {Hou},
  \citenamefont {Wang}, \citenamefont {Zheng}, \citenamefont {Tong},
  \citenamefont {Wei},\ and\ \citenamefont {Yan}}]{Dong2014}%
  \BibitemOpen
  \bibfield  {author} {\bibinfo {author} {\bibfnamefont {D.}~\bibnamefont
  {Hou}}, \bibinfo {author} {\bibfnamefont {R.}~\bibnamefont {Wang}}, \bibinfo
  {author} {\bibfnamefont {X.}~\bibnamefont {Zheng}}, \bibinfo {author}
  {\bibfnamefont {N.}~\bibnamefont {Tong}}, \bibinfo {author} {\bibfnamefont
  {J.}~\bibnamefont {Wei}},\ and\ \bibinfo {author} {\bibfnamefont
  {Y.}~\bibnamefont {Yan}},\ }\href
  {https://doi.org/10.1103/PhysRevB.90.045141} {\bibfield  {journal} {\bibinfo
  {journal} {Phys. Rev. B}\ }\textbf {\bibinfo {volume} {90}},\ \bibinfo
  {pages} {045141} (\bibinfo {year} {2014})}\BibitemShut {NoStop}%
\bibitem [{\citenamefont {Li}\ \emph {et~al.}(2012)\citenamefont {Li},
  \citenamefont {Tong}, \citenamefont {Zheng}, \citenamefont {Hou},
  \citenamefont {Wei}, \citenamefont {Hu},\ and\ \citenamefont {Yan}}]{Li2012}%
  \BibitemOpen
  \bibfield  {author} {\bibinfo {author} {\bibfnamefont {Z.}~\bibnamefont
  {Li}}, \bibinfo {author} {\bibfnamefont {N.}~\bibnamefont {Tong}}, \bibinfo
  {author} {\bibfnamefont {X.}~\bibnamefont {Zheng}}, \bibinfo {author}
  {\bibfnamefont {D.}~\bibnamefont {Hou}}, \bibinfo {author} {\bibfnamefont
  {J.}~\bibnamefont {Wei}}, \bibinfo {author} {\bibfnamefont {J.}~\bibnamefont
  {Hu}},\ and\ \bibinfo {author} {\bibfnamefont {Y.}~\bibnamefont {Yan}},\
  }\href {https://doi.org/10.1103/PhysRevLett.109.266403} {\bibfield  {journal}
  {\bibinfo  {journal} {Phys. Rev. Lett.}\ }\textbf {\bibinfo {volume} {109}},\
  \bibinfo {pages} {266403} (\bibinfo {year} {2012})}\BibitemShut {NoStop}%
\bibitem [{\citenamefont {Cirio}\ \emph {et~al.}(2022)\citenamefont {Cirio},
  \citenamefont {Kuo}, \citenamefont {Chen}, \citenamefont {Nori},\ and\
  \citenamefont {Lambert}}]{Cirio2022}%
  \BibitemOpen
  \bibfield  {author} {\bibinfo {author} {\bibfnamefont {M.}~\bibnamefont
  {Cirio}}, \bibinfo {author} {\bibfnamefont {P.-C.}\ \bibnamefont {Kuo}},
  \bibinfo {author} {\bibfnamefont {Y.-N.}\ \bibnamefont {Chen}}, \bibinfo
  {author} {\bibfnamefont {F.}~\bibnamefont {Nori}},\ and\ \bibinfo {author}
  {\bibfnamefont {N.}~\bibnamefont {Lambert}},\ }\href
  {https://doi.org/10.1103/PhysRevB.105.035121} {\bibfield  {journal} {\bibinfo
   {journal} {Phys. Rev. B}\ }\textbf {\bibinfo {volume} {105}},\ \bibinfo
  {pages} {035121} (\bibinfo {year} {2022})}\BibitemShut {NoStop}%
\bibitem [{\citenamefont {Shao}\ and\ \citenamefont {Makri}(2002)}]{Shao2002}%
  \BibitemOpen
  \bibfield  {author} {\bibinfo {author} {\bibfnamefont {J.}~\bibnamefont
  {Shao}}\ and\ \bibinfo {author} {\bibfnamefont {N.}~\bibnamefont {Makri}},\
  }\href {https://doi.org/10.1063/1.1423936} {\bibfield  {journal} {\bibinfo
  {journal} {The Journal of Chemical Physics}\ }\textbf {\bibinfo {volume}
  {116}},\ \bibinfo {pages} {507} (\bibinfo {year} {2002})}\BibitemShut
  {NoStop}%
\end{thebibliography}%

\newpage

\pagebreak
\setcounter{equation}{0}
\setcounter{figure}{0}
\setcounter{table}{0}
\setcounter{page}{1}
\setcounter{section}{0}
\makeatletter
\renewcommand{\thesection}{SM-\Roman{section}}
\renewcommand{\theequation}{S\arabic{equation}}
\renewcommand{\thefigure}{S\arabic{figure}}
\renewcommand{\bibnumfmt}[1]{[S#1]}
\renewcommand{\citenumfont}[1]{S#1}

\onecolumngrid
\begin{center}
\Large Supplemental Material for ``\inserttitle''
\end{center}

\vspace*{2em}

This supplement is divided into three parts:
\begin{itemize}
    \item[\nnref{sec:SM-additional-data}{}{}: ] Additional data
    \item[\nnref{sec:SM-IF-phase}{}{}: ] Outline of the computation of the influence phase
    \item[\nnref{sec:SM-berezin-integration}{}{}: ] Computations with the MPS-IF
\end{itemize}

\section{\label{sec:SM-additional-data}Additional data}

\subsection{Dynamics with the iteratively constructed MPS-IF}

\begin{figure}[H]
    \centering
    \includegraphics[scale=0.8]{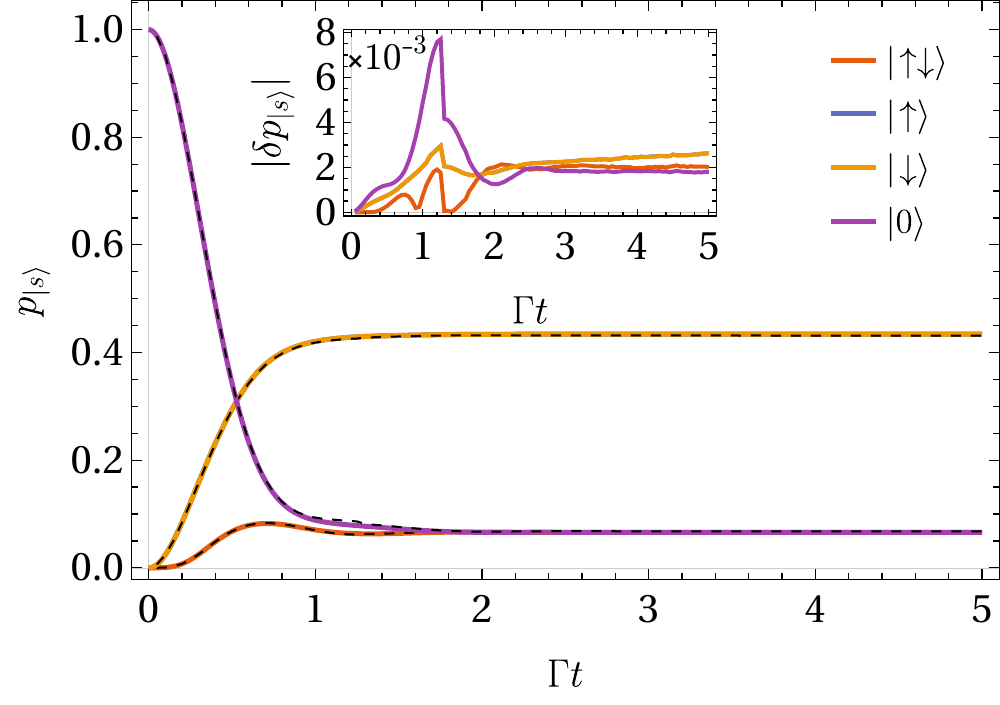}
    \caption{Impurity populations in the symmetric Anderson model with $U = 2.5\pi\Gamma$ and $\varepsilon_{\sigma} = -1.25\pi\Gamma$. The impurity is initially unoccupied. Solid colored lines are results from Ref.~\cite{Kohn2022} at temperature $\Gamma \beta = 2$, and black dashed lines are results with $D = 64$ and $\Gamma \Delta t = 0.05$, with the iteratively constructed MPS-IF. \textbf{(inset)} Absolute deviations of the populations.}
    \label{fig:pops-64-iterative-santoro-comparison}
\end{figure}

We show in \nnref{fig:pops-64-iterative-santoro-comparison}{Fig.~}{} the population dynamics in the Anderson model as generated by the iteratively constructed MPS-IF.
The data here is analogous to \nnref{fig:pops-64-santoro-comparison}{Fig.~}{} of the main text; note that the dynamics generated by the two construction methods are not identical, although the deviations are generally of the same magnitude.

\subsection{Dependence of errors on $\Delta t$ for $U = 0$}

\begin{figure}[H]
    \centering
    \includegraphics[scale=0.8]{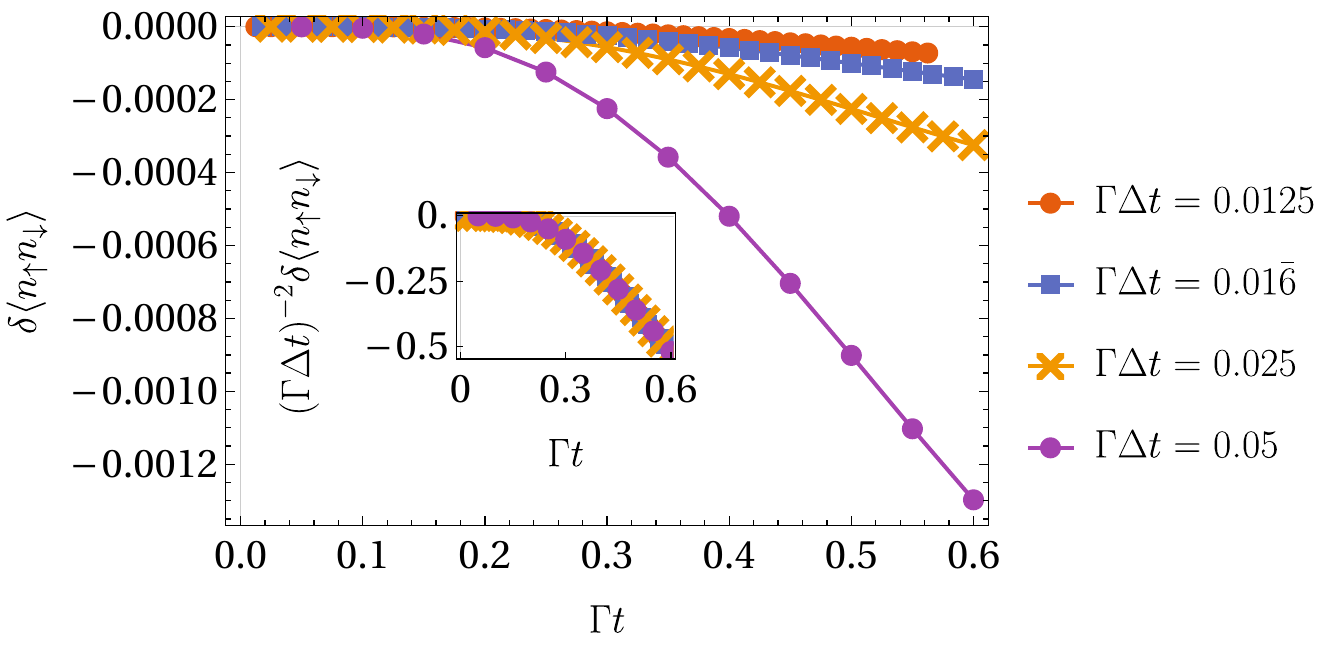}
    \caption{Deviations of the double occupancy generated by the directly constructed $D = 128$ MPS-IF from the exact values, in the $U=0$ limit and $\varepsilon_{\sigma} = -1.25\pi\Gamma$. The bath is initially at temperature $\Gamma \beta = 2$ and the impurity is initially unoccupied. \textbf{(inset)} Same data, but rescaled by the square of the timestep size.}
    \label{fig:nonint-comparison-scaling}
\end{figure}

In \nnref{fig:nonint-comparison-scaling}{Fig.~}{} we show the Trotter error made by the MPS-IF when computing the double occupancy in the noninteracting limit.
The data is generated using the $D = 128$ directly constructed MPS-IF, for which the truncation error should be negligible over the shown time range.
The collapse shown in the inset indicates that these deviations scale as $(\Delta t)^2$, as claimed in the main text.

\subsection{Entanglement in the MPS-IF}

\begin{figure}[H]
    \centering
    \includegraphics[width=\textwidth]{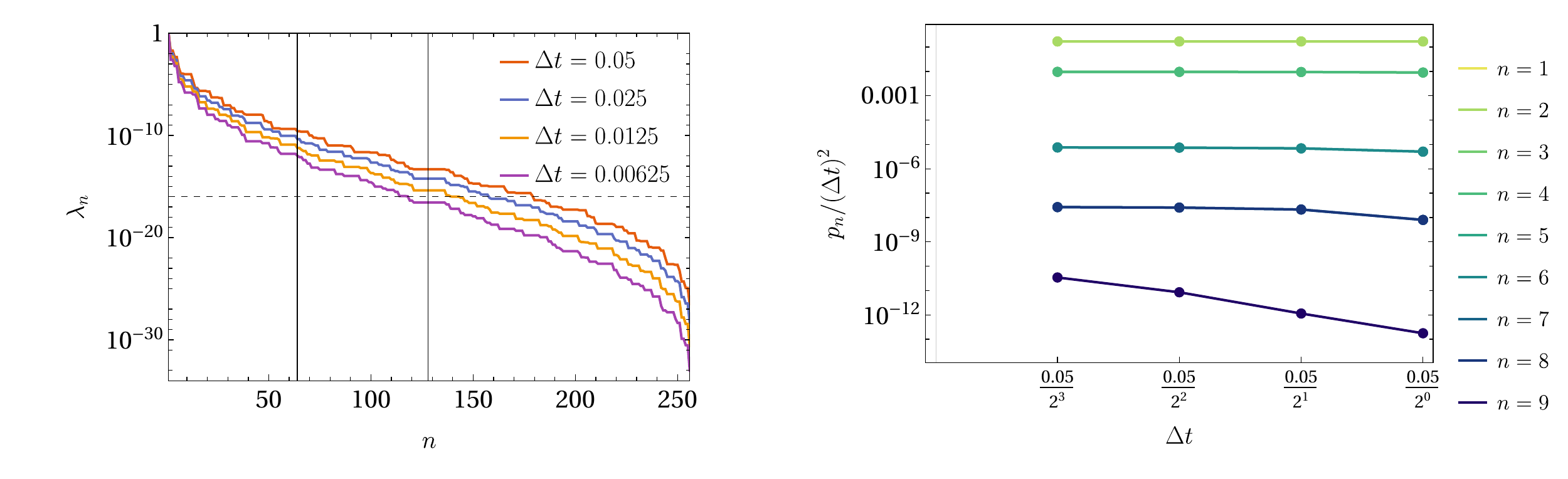}
    \caption{\textbf{(left)} Schmidt values for a bipartition of the MPS-IF into contiguous halves of the same size, for various sizes of timestep $\Delta t$. In all cases, the propagation time is fixed to $\Gamma t = 0.40$. As in the main text, we take the bath to initially be at temperature $\Gamma \beta = 2$. The vertical lines demarcate the 64th and 128th Schmidt states, while the horizontal dashed line shows the 64-bit floating point precision of $10^{-16}$.     \textbf{(right)} Single particle entanglement spectrum of the half-bipartition of $|I_N\rangle$, cf.\ Fig.~5 of Ref.~\cite{SONNER2021168677}.}
    \label{fig:schmidt-values}
\end{figure}

We make note of the entanglement properties of $|I\rangle$, which are independent of the form of the impurity Hamiltonian $\op{H}_0$.
In \nnref{fig:schmidt-values}{Fig.~}{}, we show the Schmidt values $\lambda_n$ of the MPS-IF for a bipartition of $|I\rangle$ for various sizes of timestep $\Delta t$.
We observe that the Schmidt values decay superpolynomially but subexponentially for the range of $n$ that we have used in our calculations. 
We also observe that the single-particle entanglement spectrum, $p_n$, scales as $(\Delta t)^2$ for four different timesteps in \nnref{fig:schmidt-values}{Fig.~}{}.
It is consistent with the result from the integrable quantum spin chain model \cite{SONNER2021168677}, which leads to $(\Delta t)^2$ scaling of the entanglement entropy.

\newpage 

\section{\label{sec:SM-IF-phase}Computation of the matrix $\mat{G}$}

\subsection{Derivation of the influence functional}

For the single impurity problem, we can simplify the calculation of the IF by separating the up-spin bath fermions from the down-spin ones.
In the absence of spin-dependent bias, the influence functional for the bath will turn out to factorize into the product of identical IFs for the up-spin fermions and the down-spin fermions.

Using the shorthand notation $|s\rangle\langle s'| \equiv |s_{\uparrow}, s_{\downarrow}\rangle\langle s'_{\uparrow}, s'_{\downarrow}| \otimes \op{I}_{B}$, the impurity density matrix is propagated as,
\begin{align}
  \label{eq:path-integral}
  \langle s^+_N | \op{\widetilde{\rho}}(t) | s^-_N\rangle &\approx \sum\limits_{\{s^{\pm}_i\}} \Tr\Bigg[ \dop{s^-_N}{s^+_N}{\op{I}_B} \op{U}_0 \left(\frac{\delta t}{2} \right) \\
  &\quad \times \dop{s^+_{N^-}}{s^+_{N^-}}{\op{I}_B} \op{U}_1(\delta t) \dop{s^+_{N-1^+}}{s^+_{N-1^+}}{\op{I}_B} \op{U}_0(\delta t) \cdots \dop{s^+_{0^+}}{s^+_{0^+}}{\op{I}_B} \op{U}_0\left( \frac{\delta t}{2} \right) \nonumber \\
  &\quad \times \dop{s^+_0}{s^+_0}{\op{I}_B} \op{\widetilde{\rho}}(0) \otimes \op{\widetilde{\rho}}_B \dop{s^-_0}{s^-_0}{\op{I}_B} \nonumber \\
  &\quad \times \op{U}_0\left( \frac{-\delta t}{2} \right) \dop{s^-_{0^+}}{s^-_{0^+}}{\op{I}_B} \cdots \op{U}_0(-\delta t) \dop{s^-_{N-1^+}}{s^-_{N-1^+}}{\op{I}_B} \op{U}_1(-\delta t) \dop{s^-_{N^-}}{s^-_{N^-}}{\op{I}_B} \nonumber \\
  &\quad \times \op{U}_0 \left( \frac{- \delta t}{2} \right) \Bigg]. \nonumber
\end{align}

$\op{U}_0$ ($\op{U}_1$) is the time evolution operator from $\op{H}_0$ ($\op{H}_1$). The pure system propagators can be factored out of \nnref{eq:path-integral}{Eq.~(}{)}, and one is left with the influence functional,
\begin{align}
\label{eq:unsplit-IF}
\begin{split}
  I(\{s^{\pm}\}) &\approx \Tr \Bigg[ \dop{s^-_{N^-}}{s^+_{N^-}}{\op{I}_{\sigma,B}} \op{U}_1(\delta t) \dop{s^+_{N-1^+}}{s^+_{N-1^-}}{\op{I}_{B}} \\
  &\qquad\quad \times \cdots \times \dop{s^+_{1^+}}{s^+_{1^-}}{\op{I}_{B}} \op{U}_1(\delta t) \Big\{ |s^+_{0^+} \rangle\langle s^-_{0^+}| \otimes \op{\widetilde{\rho}}_B \Big\} \op{U}_1(-\delta t) \dop{s^-_{1^-}}{s^-_{1^+}}{\op{I}_{B}} \\
  &\qquad\quad \times \cdots \times \dop{s^-_{N-1^-}}{s^-_{N-1^+}}{\op{I}_{B}} \op{U}_1(-\delta t) \Bigg].
\end{split}
\end{align}
We denote this as $I$, the influence functional in the computational basis defined by spin states.
Note that in the influence functional, the two spin sectors do not mix with each other and therefore the IF factorizes into two parts.
Each trace can be evaluated separately by first undoing the Jordan-Wigner and chain-mapping transformations.
The spin states map back to fermions as,
\begin{align}
\label{eq:spin-fermion-map}
\begin{split}
  |+1\rangle\langle -1| &\equiv \op{d}^{\dagger}\phantom{\op{d}}, \hspace*{1in} |-1\rangle\langle +1| \equiv \op{d}, \\
  |+1\rangle\langle +1| &\equiv \op{d}^{\dagger}\op{d}, \hspace*{1in} |-1\rangle\langle -1| \equiv \op{d}\op{d}^{\dagger},
\end{split}
\end{align}
and the propagator $\op{U}_1$ is bilinear,
\begin{align*}
\op{U}_1(\delta t) &= \prod_{\sigma} \exp \left[ -i \delta t \left( \sum_{k} \varepsilon_k \op{c}_{k,\sigma}^{\dagger} \op{c}_{k,\sigma} + V_k \op{d}_{\sigma}^{\dagger} \op{c}_{k,\sigma} + V^*_k \op{c}_{k,\sigma}^{\dagger} \op{d}_{\sigma} \right) \right].
\end{align*}
Note that the two spin sectors have now decoupled and the influence functional factorizes.
We will therefore focus our discussion only on the influence functional for $\sigma = \downarrow$.

We can cast $I$ in a way reminiscent of the spin-boson case by working in the basis of coherent states, which is also essential for the construction of an influence functional in tensor network form.
Using the following notation for compactness,
\begin{align*}
\op{1}(\psi, \chi) &= \Bigg( \int d\chi d\psi \, e^{\psi \chi} |\psi \rangle\langle \chi| \Bigg),
\end{align*}
we insert resolutions of the identity into \nnref{eq:unsplit-IF}{Eq.~(}{)} to obtain
\begin{align}
\label{eq:grassmann-IF}
  I(\{s^{\pm}\}) &\approx \Tr \Bigg[ \roi{\gB{\eta}{N^-}}{\gBc{\eta}{N^-}} \dop{s^-_{N^-}}{s^+_{N^-}}{\op{I}_{\sigma,B}} \roi{\gF{\eta}{N^-}}{\gFc{\eta}{N^-}} \op{U}_1(\delta t) \roi{\gF{\eta}{N-1^+}}{\gFc{\eta}{N-1^+}} \dop{s^+_{N-1^+}}{s^+_{N-1^-}}{\op{I}_{B}} \nonumber \\
  &\qquad\quad \times \cdots \times \dop{s^+_{1^+}}{s^+_{1^-}}{\op{I}_{B}} \roi{\gF{\eta}{1^-}}{\gFc{\eta}{1^-}} \op{U}_1(\delta t) \roi{\gF{\eta}{0^+}}{\gFc{\eta}{0^+}} \Big\{ |s^+_{0^+} \rangle\langle s^-_{0^+}| \otimes \op{\widetilde{\rho}}_B \Big\} \nonumber \\
                   &\qquad\quad \times \roi{\gB{\eta}{0^+}}{\gBc{\eta}{0^+}} \op{U}_1(-\delta t) \roi{\gB{\eta}{1^-}}{\gBc{\eta}{1^-}} \dop{s^-_{1^-}}{s^-_{1^+}}{\op{I}_{B}} \times \cdots \nonumber \\
  &\quad\qquad \times \dop{s^-_{N-1^-}}{s^-_{N-1^+}}{\op{I}_{B}} \roi{\gB{\eta}{N-1^+}}{\gBc{\eta}{N-1^+}} \op{U}_1(-\delta t) \Bigg].
\end{align}
It is useful to note the matrix elements of the mapping \nnref{eq:spin-fermion-map}{Eq.~(}{)},
\begin{align}
\label{eq:c-basis-kernels}
\langle \eta_1 | \dop{s_1}{s_2}{\op{I}_{\sigma,B}} |\eta_2 \rangle &= \left( \frac{1 - s_1}{2} + \frac{1 + s_1}{2} \eta_1 \right) \left( \frac{1 - s_2}{2} + \frac{1 + s_2}{2} \eta_2 \right).
\end{align}
These matrix elements allow us to convert the influence functional from a Grassmann representation to a commuting basis, e.g.\ the four-state basis of the single Anderson impurity.
Factoring these out of the trace, which can be done so long as the ordering of the terms is preserved, leaves us with Grassmann integrals involving only the propagators $\op{U}_1$ in the fermion coherent state basis.
The matrix elements of these propagators can be evaluated as,
\begin{align}
\begin{split}
  \langle \gFc{\eta}{n^-}, \gFc{\xi}{n^-} | \op{U}_1(\delta t) | \gF{\eta}{n-1^+}, \gF{\xi}{n-1^+} \rangle &= \exp \left[ \begin{pmatrix} \gFc{\eta}{n^-} & \gFc{\xi}{n^-} \end{pmatrix} \cdot \mat{U}(\delta t) \cdot \begin{pmatrix} \gFc{\eta}{n^-} \\ \gF{\xi}{n-1^+} \end{pmatrix} \right], \\
  \mat{U}(\delta t) &= \exp \left[ -i \delta t \left( \sum_k V_k |0\rangle\langle k| + V_k^{*} |k\rangle\langle 0| + E_k |k\rangle\langle k| \right) \right].
\end{split}
\end{align}
$|0\rangle$ and $|k\rangle$ are one-particle basis for impurity and bath degrees of freedom, respectively.

In all, the IF corresponding to the specified Trotterized dynamics can be exactly evaluated to be
\begin{align}
\begin{split}
  I &= \exp \Bigg[ F(\Delta t) \sum_{m=0}^{N-1} \gFc{\eta}{m+1^-} \gF{\eta}{m^+} - F^{*}(\Delta t) \sum_{m=0}^{N-1} \gB{\eta}{m+1^-} \gBc{\eta}{m^+} \Bigg] \\
  &\quad \times \frac{Z_N}{Z} \exp \Bigg[ \sum\limits_{m=1}^{N-1} \sum\limits_{n=1}^m \gFc{\eta}{m+1^-} \gF{\eta}{n-1^+} G^{++}_{>}(m,n) + \sum\limits_{m=1}^{N} \sum\limits_{n=0}^{m-1} \gF{\eta}{m-1^+} \gFc{\eta}{n+1^-} G^{++}_{<}(m,n) \\
  &\hspace*{1in} - \sum\limits_{m=1}^{N-1} \sum\limits_{n=1}^m \gB{\eta}{m+1^-} \gBc{\eta}{n-1^+} \Big( G^{++}_{>}(m,n)\Big)^{*} - \sum\limits_{m=1}^{N} \sum\limits_{n=0}^{m-1} \gBc{\eta}{m-1^+} \gB{\eta}{n+1^-} \Big( G^{++}_{<}(m,n) \Big)^{*} \\
  &\hspace*{1in} + \sum\limits_{m=1}^N \sum\limits_{n=1}^m \gBc{\eta}{m-1^+} \gF{\eta}{n-1^+} G_{-}(m,n) + \sum\limits_{m=0}^{N-1} \sum\limits_{n=0}^{m} \gB{\eta}{m+1^-} \gFc{\eta}{n+1^-} G_{+}(m,n) \\
  &\hspace*{1in} - \sum\limits_{m=1}^N \sum\limits_{n=1}^{m-1} \gF{\eta}{m-1^+} \gBc{\eta}{n-1^+} \Big( G_{-}(m,n) \Big)^{*} - \sum\limits_{m=0}^{N-1} \sum\limits_{n=0}^{m-1} \gFc{\eta}{m+1^-} \gB{\eta}{n+1^-} \Big( G_{+}(m,n) \Big)^{*} \Bigg].
\end{split}
\end{align}
Here, $Z = \Tr_B \exp(-\beta \op{H}_B)$ corresponds to the partition function of the initially equilibrated non-interacting bath, and $Z_N$ corresponds to the normalization factor for bath evolving under non-unitary dynamics~\cite{Thoenniss2022}.

\subsection{Constructing $\mat{G}$}
The splitting of the Hamiltonian used in the influence functional formalism, which is essentially an expansion in terms of the interaction strength $U$, leads to a Trotterization scheme in which the system is alternately propagated by a non-interacting Hamiltonian or by the Hamiltonian of interactions between different components of the impurity.
Thus the temporal correlations appearing in the influence functional are completely characterized by the solution of the non-interacting problem, which can be numerically computed from the hybridization function $\Delta_{\sigma}(t) = \int d\epsilon \, |V_{\sigma}(\epsilon)|^2 \rho(\epsilon) e^{-i \epsilon t}$.

Note that there are different ways to split the Hamiltonian so as to leave a numerically solvable non-interacting problem.
For example, terms that are quadratic in the impurity degrees of freedom such as the onsite energy can either be grouped into $\op{H}_0$ or $\op{H}_1$ as defined Eq.~(1) of the main text.

In this section we calculate the matrix elements appearing in the definition of influence functional,
\begin{align*}
\langle \gFc{\eta}{}, \gFc{\xi}{} | \exp \left[ -i \Delta t \sum\limits_k \Big( V_k \op{d}^{\dagger} \op{c}_k + \hc \Big) + E_k \op{c}^{\dagger}_k \op{c}_k \right] | \gF{\eta}{}, \gF{\xi}{} \rangle
\end{align*}
which can be evaluated as
\begin{align}
\label{eq:fermion-matelem}
\exp \Bigg[ 
\begin{pmatrix}
\gFc{\eta}{} & \gFc{\xi}{}
\end{pmatrix} \cdot
\underbrace{\exp \left\{ -i \Delta t \begin{pmatrix}
  0 & \vec{V}^{\top} \\
  \vec{V}^{*} & \mat{h}_B
\end{pmatrix} \right\}}_{\equiv \mat{U}(\Delta t)} \cdot
\begin{pmatrix}
  \gF{\eta}{} \\
  \gF{\xi}{}
\end{pmatrix}
 \Bigg].
\end{align}

The matrix $\mat{U}(\Delta t)$ can be explicitly computed (see next section).
The matrix element \nnref{eq:fermion-matelem}{Eq.~(}{)} can be split suggestively as
\begin{align*}
\begin{split}
  \langle \gFc{\eta}{}, \gFc{\xi}{} | \op{U}_1(\Delta t) | \gF{\eta}{}, \gF{\xi}{} \rangle &= \exp \left[ U_{00}(\Delta t) \gFc{\eta}{} \gF{\eta}{} \right] \\
  &\quad \times \exp \Big[ \gFc{\eta}{} \left( U_{0k} \gF{\xi}{k} \right) \Big] \exp \Big[ \gFc{\xi}{k} U_{kk'} \gF{\xi}{k'} \Big] \exp \Big[ \left( \gFc{\xi}{k} U_{k0} \right) \gF{\eta}{} \Big]
\end{split} \\
\begin{split}
  &= \exp \left[ U_{00}(\Delta t) \gFc{\eta}{} \gF{\eta}{} \right] \\
  &\quad \times \langle \gFc{\xi}{} | \exp \Big[ \gFc{\eta}{} \left( U_{0k} \op{c}_k \right) \Big] \exp \Big[ \op{c}_k^{\dagger} (\log \mat{U}_{B})_{kk'} \op{c}_{k'} \Big] \exp \Big[ \left( \op{c}_k^{\dagger} U_{k0} \right) \gF{\eta}{} \Big] |\gF{\xi}{}\rangle,
\end{split}
\end{align*}
where $\mat{U}_B$ consists of only the subblock of $\mat{U}(\Delta t)$ describing the bath degrees of freedom, i.e.\ $(\mat{U}_B)_{kk'} = [\mat{U}(\Delta t)]_{kk'}$.
Observe that $\log \mat{U}_B$ is generally not skew-Hermitian.
We will find that $\mat{U}_B$ is responsible for generating the time-evolution relevant for the bath correlation functions appearing in the influence functional.

\subsubsection{Computation of $U_{\alpha,\alpha'}(t)$}
The matrix elements of \[ \mat{U}(t) = \exp \left\{ -i \Delta t \begin{pmatrix}
  \varepsilon_{\sigma} & \vec{V}^{\top} \\
  \vec{V}^{*} & \mat{h}_B
\end{pmatrix} \right\}, \]
can be straightforwardly computed, as this unitary matrix simply describes the evolution of a distinguished state ``0'' coupled to a continuum.
The onsite energy $\varepsilon_{\sigma}$ is set to zero depending on the splitting chosen.
The evolution is thus governed by
\begin{align*}
\frac{d}{dt} \mat{U}(t) &= -i \left( \varepsilon_{\sigma} |0\rangle\langle 0| + \sum_k V_k |0\rangle\langle k| + V_k^{*} |k\rangle\langle 0| + E_k |k\rangle\langle k| \right) \mat{U}(t).
\end{align*}
Taking matrix elements of this equation gives the following system of equations,
\begin{align*}
  \frac{d}{dt} U_{00}(t) &= -i \left( \varepsilon U_{00}(t) + \sum_k V_k U_{k0}(t) \right), && \frac{d}{dt} U_{0k'}(t) = -i \left( \varepsilon U_{0k'}(t) + \sum_k V_k U_{kk'}(t) \right), \\
  \frac{d}{dt} U_{k0}(t) &= -i \Bigg( V_k^{*} U_{00}(t) + E_k U_{k0}(t) \Bigg), && \frac{d}{dt} U_{kk'}(t) = -i \Bigg( V_k^{*} U_{0k'}(t) + E_k U_{kk'}(t) \Bigg).
\end{align*}
In terms of the function $F(t)$ whose Laplace transform is \[ \widetilde{F}(z) = \left( z + i \varepsilon_{\sigma} + \sum_k \frac{|V_k|^2}{z +i E_k} \right)^{-1}, \]
the above matrix elements in the time-domain are
\begin{align}
  U_{00}(t) &= F(t), && U_{0k'}(t) = -i V_{k'} \int\limits_0^t d\tau \, e^{-i E_{k'} (t - \tau)} F(\tau) \\
  U_{k0}(t) &= -i V_{k}^{*} \int\limits_0^t d\tau \, e^{-i E_k (t - \tau)} F(\tau), && U_{kk'}(t) = e^{-i E_k t} \delta_{k, k'} - i V_k^{*} V_{k'} \int\limits_0^t d\tau \frac{e^{-i E_k \tau} - e^{-i E_{k'} \tau}}{E_k - E_{k'}} F(t - \tau). \nonumber
\end{align}

As we will see, the essential quantity for specifying the matrix elements is the fidelity $F(t)$.
This is computed in practice by solving a Volterra integrodifferential equation of second order with $F(0) = 1$,
\begin{align}
  \frac{d}{dt} F(t) &= - i \varepsilon_{\sigma} \, F(t)
                      - \int\limits_0^t d\tau \, \left( \int dk \, \left| V(k) \right|^2 e^{-i E(k) (t - \tau)} \right) F(\tau) \nonumber \\
  &\equiv - i \varepsilon_{\sigma} \, F(t) - \int\limits_0^t d\tau \, \Delta(t-\tau) F(\tau), \label{eq:SM-1-volterra}
\end{align}
where $\Delta(t)$ is the hybridization function.
Note that the foregoing discussion is easily generalizable to the multi-impurity case.

Observe that the $U=0$ impurity dynamics can be easily found from $F(t)$.
For example, the evolution of the occupation $\langle \op{n}_{\sigma}(t) \rangle$ starting from $\langle \op{n}_{\sigma}(0) \rangle = 0$ is given by
\begin{align}
    \langle \op{n}_{\sigma}(t) \rangle &= \int dk \, \frac{|V(k)|^2}{1 + \exp(\beta E(k))} \left| \int\limits_{0}^{t} d\tau_1 \, e^{-i E(k) (t-\tau_1)} F(\tau_1) \right|^2. \label{eq:SM-1-nonint-pop}
\end{align}
In the main text, we use \nnref{eq:SM-1-volterra}{Eqs.~(}{)} and \nnref{eq:SM-1-nonint-pop}{(}{)} to compute the exact non-interacting double occupancy.

\subsubsection{Evaluation of the matrix $\mat{G}$}

We now introduce the the notation for fermions evolving on, respectively, the forwards and the backwards contours,
\begin{align}
  \op{c}_i^{(\dagger)}[-T] &= \exp \Big( T \op{c}^{\dagger} \cdot (\log \mat{U}_B) \cdot \op{c} \Big) \op{c}_i^{(\dagger)} \exp \Big( -T \op{c}^{\dagger} \cdot (\log \mat{U}_B) \cdot \op{c} \Big) \\
  \ubar{\op{c}}_i^{(\dagger)}[-T] &= \exp \Big( -T \op{c}^{\dagger} \cdot (\log \mat{U}_B)^{\dagger} \cdot \op{c} \Big) \op{c}_i^{(\dagger)} \exp \Big( T \op{c}^{\dagger} \cdot (\log \mat{U}_B)^{\dagger} \cdot \op{c} \Big).
\end{align}
Tracing out the bath gives
\begin{align}
\begin{split}
  I &= \exp \Bigg[ F(\Delta t) \sum_{m=1}^N \gFc{\eta}{m^-} \gF{\eta}{m-1^+} - F^{*}(\Delta t) \sum_{m=1}^N \gB{\eta}{m^-} \gBc{\eta}{m-1^+} \Bigg] \\
  &\quad \times \frac{Z_N}{Z} \exp \Bigg[ \sum\limits_{m=1}^{N-1} \sum\limits_{n=1}^m \gFc{\eta}{m+1^-} \gF{\eta}{n-1^+} G^{++}_{>}(m,n) + \sum\limits_{m=1}^{N} \sum\limits_{n=0}^{m-1} \gF{\eta}{m-1^+} \gFc{\eta}{n+1^-} G^{++}_{<}(m,n) \\
  &\hspace*{1in} - \sum\limits_{m=1}^{N-1} \sum\limits_{n=1}^m \gB{\eta}{m+1^-} \gBc{\eta}{n-1^+} G^{--}_{>}(m,n) - \sum\limits_{m=1}^{N} \sum\limits_{n=0}^{m-1} \gBc{\eta}{m-1^+} \gB{\eta}{n+1^-} G^{--}_{<}(m,n) \\
  &\hspace*{1in} - \sum\limits_{m=1}^N \sum\limits_{n=1}^N \gBc{\eta}{m-1^+} \gF{\eta}{n-1^+} G_{-}(m,n) - \sum\limits_{m=0}^{N-1} \sum\limits_{n=0}^{N-1} \gB{\eta}{m+1^-} \gFc{\eta}{n+1^-} G_{+}(m,n) \Bigg],
\end{split}
\end{align}
where we have defined each elements
\begin{align}
\begin{split}
  G^{++}_{>}(m, n) &= U_{0,i} \, \Tr \Bigg[ \op{c}_i[-(N-m)] \, \op{c}_j^{\dagger}[-(N-n)] \, \op{\rho}_B^{(N)} \Bigg] U_{j,0}, \\
  G^{++}_{<}(m, n) &= U_{i,0} \, \Tr \Bigg[ \op{c}_i^{\dagger}[-(N-m)] \, \op{c}_j[-(N-n)] \, \op{\rho}_B^{(N)} \Bigg] U_{0,j},
\end{split} \\
\begin{split}
  G^{--}_{>}(m, n) &= U_{i,0}^{*} \, \Tr \Bigg[ \ubar{\op{c}}_i[-(N-n)] \, \ubar{\op{c}}_j^{\dagger}[-(N-m)] \, \op{\rho}_B^{(N)} \Bigg] U_{0,j}^{*}, \\
  G^{--}_{<}(m, n) &= U_{0,i}^{*} \, \Tr \Bigg[ \ubar{\op{c}}_i^{\dagger}[-(N-n)] \, \ubar{\op{c}}_j[-(N-m)] \, \op{\rho}_B^{(N)} \Bigg] U_{j,0}^{*},
\end{split} \\
\begin{split}
  G_-(m, n) &= U_{i,0}^{*} \, \Tr \Bigg[ \ubar{\op{c}}_i[-(N-m)] \, \op{c}_j^{\dagger}[-(N-n)] \, \op{\rho}_B^{(N)} \Bigg] U_{j,0}, \\
  G_+(m, n) &= U_{0,i}^{*} \, \Tr \Bigg[ \ubar{\op{c}}_i^{\dagger}[-(N-m)] \, \op{c}_j[-(N-n)] \, \op{\rho}_B^{(N)} \Bigg] U_{0,j}.
\end{split}
\end{align}
Here we have definitions
\begin{align}
  \mat{I}_B^{(n)} &= \Big( \mat{U}_B^{\dagger} \Big)^n \Big( \mat{U}_B \Big)^n, \\
  \op{I}_B^{(n)} &= \exp \Bigg[ \op{c}_i^{\dagger} \Big( \log \mat{I}_B^{(n)} \Big)_{ij} \op{c}_j \Bigg], \\
  \op{\rho}_B^{(n)} &= \frac{1}{Z_n} \exp \left[ \op{c}_i^{\dagger} \Bigg( \log \Bigg\{ \Big( \mat{U}_B \Big)^n \Big( e^{-\beta \mat{h}_B} \Big) \Big( \mat{U}_B^{\dagger} \Big)^n \Bigg\} \Bigg)_{ij} \op{c}_j \right],
\end{align}
with $Z_n$ such that $\Tr_B \op{\rho}_B^{(n)} = 1$, and $Z \equiv Z_0$.

These correlation functions have several notable properties.
First, they are not time-translationally invariant, e.g.\ $G^{++}_{>}(m, n) \neq G^{++}_{>}(m-n, 0)$.
Second, they explicitly depend on the total propagation time, $N\Delta t$.
One sees that
\begin{align*}
\begin{pmatrix}
\op{c}_1[-T] \\ 
\vdots \\
\op{c}_M[-T]
\end{pmatrix} = 
\left( \mat{U}_B\right)^{-T} \cdot \begin{pmatrix}
\op{c}_1 \\ 
\vdots \\
\op{c}_M
\end{pmatrix}
&& \text{and} \qquad
\begin{pmatrix}
\ubar{\op{c}}_1[-T] \\ 
\vdots \\
\ubar{\op{c}}_M[-T]
\end{pmatrix} = 
\left( \mat{U}_B^{\dagger}\right)^{T} \cdot \begin{pmatrix}
\op{c}_1 \\ 
\vdots \\
\op{c}_M
\end{pmatrix},
\end{align*}
with the remaining relations following from Hermitian conjugation of the above two cases.
From this, it is clear that the expected relationships for the correlation functions are present,
\begin{align}
  G^{++}_{>}(m,n) &= \Big( G^{--}_{>}(m,n) \Big)^{*}, && G^{++}_{<}(m,n) = \Big( G^{--}_{<}(m,n) \Big)^{*}, \\
  G_{-}(m,n) &= \Big( G_{-}(n,m) \Big)^{*}, && \,\,\,\, G_{+}(m,n) = \Big( G_{+}(n,m) \Big)^{*}.
\end{align}

It is not trivial to invert the nonunitary matrix $\mat{U}_B$, so it is advantageous to express the correlation functions $G(m,n)$ only in terms of $\mat{U}_B$ raised to nonnegative integer powers.
Doing so yields the following representations of the correlation functions (summations are implied)
\begin{align}
\begin{split}
  G^{++}_{>}(m, n) &= \frac{Z_n}{Z_N} \Bigg\{ U_{0,k} \left( \mat{U}_B^{m-n} \right)_{k,i} \Bigg\} \,\, \Tr \Big[ \op{c}_i \, \op{c}_j^{\dagger} \, \op{\rho}_B^{(n)} \op{I}_B^{(N-n)} \Big] \,\, U_{j,0} \\
  G^{++}_{<}(m, n) &= \frac{Z_n}{Z_N} \Bigg\{ \left( \mat{U}_B^{-(m-n)} \right)_{i,k} U_{k,0} \Bigg\} \,\, \Tr \Big[ \op{c}_i^{\dagger} \, \op{c}_j \, \op{\rho}_B^{(n)} \op{I}_B^{(N-n)} \Big] \,\, U_{0,j}
\end{split} \\
\begin{split}
  G_-(m, n) &= \Bigg\{ \left( \mat{U}_B^{N-m} \right)_{i,k} U_{k,0} \Bigg\}^{*} \,\, \Tr \Big[ \op{c}_i \, \op{c}_j^{\dagger} \, \op{\rho}_B^{(N)} \Big] \,\, \Bigg\{ \left( \mat{U}_B^{N-n} \right)_{j,k'} U_{k',0} \Bigg\} \\
  G_+(m, n) &= \frac{Z}{Z_N} \Bigg\{ U_{0,k} \left( \mat{U}_B^m \right)_{k,i} \Bigg\}^{*} \,\, \Tr \Big[ \op{c}_i^{\dagger} \, \op{I}_B^{(N)} \, \op{c}_j \, \op{\rho}_B \Big] \,\, \Bigg\{ U_{0,k} \left( \mat{U}_B^n \right)_{k,j} \Bigg\}.
\end{split}
\end{align}
The quantities $\sum_k U_{0,k} \left( \mat{U}_B^n \right)_{k,i}$ and $\sum_k \left( \mat{U}_B^n \right)_{i,k} U_{k,0}$ appearing in the correlation functions are straightforwardly computable and can be constructed recursively, so long as $F(t) \equiv U_{0,0}(t)$ and $U_{0,k}(t)$ and $U_{k,0}(t)$ are known.
We sketch this in a later section.

In the end, note that both $\op{\rho}_B^{(n)}$ and $\op{I}_B^{(n)}$ constitute quadratic operators, so the bath traces in the correlation functions can be computed exactly from
\begin{align}
  \Tr \Big[ \op{c}_i \, \op{c}_j^{\dagger} \, e^{\op{c}^{\dagger} \cdot \mat{A} \cdot \op{c}} \Big] &= \Big( \Tr \, e^{\op{c}^{\dagger} \cdot \mat{A} \cdot \op{c}} \Big) \left( \frac{1}{\mat{I} + e^{\mat{A}}} \right)_{i,j}, \\
  \Tr \Big[ \op{c}^{\dagger}_i \, \op{c}_j \, e^{\op{c}^{\dagger} \cdot \mat{A} \cdot \op{c}} \Big] &= \Big( \Tr \, e^{\op{c}^{\dagger} \cdot \mat{A} \cdot \op{c}} \Big) \left( \frac{1}{\mat{I} + e^{\mat{A}}} e^{\mat{A}} \right)_{j,i}.
\end{align}
Thus (formally) we have
\begin{align}
\begin{split}
  G^{++}_{>}(m, n) &= \Bigg\{ U_{0,k} \left( \mat{U}_B^{m-n} \right)_{k,i} \Bigg\} \left( \frac{1}{\mat{I} + \mat{U}_B^n e^{-\beta \mat{h}_B} \left( \mat{U}_B^N \right)^{\dagger} \mat{U}_B^{N-n}} \right)_{i,j} U_{j,0}, \\
  G^{++}_{<}(m, n) &= U_{0,j} \left( \mat{I} - \frac{1}{\mat{I} + \mat{U}_B^n e^{-\beta \mat{h}_B} \left( \mat{U}_B^N \right)^{\dagger} \mat{U}_B^{N-n}} \right)_{j,i} \Bigg\{ \left( \mat{U}_B^{-(m-n)} \right)_{i,k} U_{k,0} \Bigg\}, 
\end{split} \\
\begin{split}
  G_-(m, n) &= \Bigg\{ \left( \mat{U}_B^{N-m} \right)_{i,k} U_{k,0} \Bigg\}^{*} \left( \frac{1}{\mat{I} + \mat{U}_B^N e^{-\beta \mat{h}_B} \left( \mat{U}_B^N \right)^{\dagger}} \right)_{i,j} \Bigg\{ \left( \mat{U}_B^{N-n} \right)_{j,k'} U_{k',0} \Bigg\}, \\
  G_+(m, n) &= \Bigg\{ U_{0,k} \left( \mat{U}_B^n \right)_{k,j} \Bigg\} \left( e^{-\beta \mat{h}_B} \frac{1}{\mat{I} + \left( \mat{U}_B^N \right)^{\dagger} \mat{U}_B^N e^{-\beta \mat{h}_B}} \right)_{j,i} \Bigg\{ U_{0,k} \left( \mat{U}_B^m \right)_{k,i} \Bigg\}^{*}.
\end{split}
\end{align}
Note that the troublesome correlation function $G^{++}_{<}(m,n)$ contains the bath-only part of the unitary evolution raised to a negative power, $\mat{U}_B^{-(m-n)}$.
To get around the need to invert arbitrarily large matrices, we note that
\begin{align*}
  \mat{U}_B^{-m} &= \mat{U}_B^{-m} \left( \mat{U}_B^{\dagger} \right)^{-m} \left( \mat{U}_B^{\dagger} \right)^{m} \\
  &= \Big[ \mat{I}_B^{(m)} \Big]^{-1} \left( \mat{U}_B^{\dagger} \right)^{m}.
\end{align*}
Observe that we have traded the explicit inversion of $\mat{U}_B$ for the explicit inversion of $\mat{I}_B^{(m)}$.
The application of $\mat{U}_B^{\dagger}$ on $U_{k,0}$ can be computed by noting that,
\begin{align*}
  \sum_k \left( \mat{U}_B^{\dagger} \right)_{i,k} U_{k,0} &= -F(\Delta t) \mat{U}_{0,i}^{*} \quad \Longrightarrow \quad \sum_k \left( (\mat{U}_B^{\dagger})^n \right)_{i,k} U_{k,0} = -F(\Delta t) \Bigg[ \sum_k U_{0,k} \left( \mat{U}_B^{n-1} \right)_{k,i} \Bigg]^{*}.
\end{align*}

\subsubsection{Evaluation of correlation functions}
Recall that the full unitary matrix $\mat{U}(t)$ has the block structure
\begin{align*}
  \mat{U}(t) &= \exp \left[ -i t \begin{pmatrix} 0 & \vline & \vec{V}^{\intercal} & \\ \hline & \vline & & \\ \vec{V}^{*} & \vline & \mat{h}_B & \\ & \vline & & \end{pmatrix} \right] \quad \Longrightarrow \quad \left[ \mat{U}(t) \right]_{k,k'} \equiv \begin{pmatrix} F(t) & \vline & U_{0,k'}(t) & \\ \hline & \vline & & \\ U_{k,0}(t) & \vline & \left(\mat{U}_B\right)_{k,k'}(t) & \\ & \vline & & \end{pmatrix}.
\end{align*}
From this structure it immediately follows that the application of $\mat{U}_B(\Delta t)$ on $U_{0,k}(\Delta t)$ is related to $U_{0,k}(2 \Delta t)$,
\begin{align*}
  \left[ \mat{U}(2t) \right]_{k,k'} &= \begin{pmatrix} F(2t) & \vline & U_{0,k'}(2t) & \\ \hline & \vline & & \\ U_{k,0}(2t) & \vline & \left(\mat{U}_B\right)_{k,k'}(2t) & \\ & \vline & & \end{pmatrix} \\
                                    &= \begin{pmatrix} F(t) & \vline & U_{0,k'}(t) & \\ \hline & \vline & & \\ U_{k,0}(t) & \vline & \left(\mat{U}_B\right)_{k,k'}(t) & \\ & \vline & & \end{pmatrix} \begin{pmatrix} F(t) & \vline & U_{0,k'}(t) & \\ \hline & \vline & & \\ U_{k,0}(t) & \vline & \left(\mat{U}_B\right)_{k,k'}(t) & \\ & \vline & & \end{pmatrix} \\
  &= \begin{pmatrix} {} & \vline & F(t) U_{0,k'}(t) + \sum_i U_{0,k}(t) \left( \mat{U}_B(t) \right)_{k,k'}  & \\ \hline & \vline & & \\ {} & \vline & {} & \\ & \vline & & \end{pmatrix}.
\end{align*}
Let us introduce the notation
\begin{align*}
  F^{(m,n)}_k &= \sum_{k'} U_{0, k'}(n \Delta t) \left( \mat{U}_B \right)_{k',k}(m \Delta t).
\end{align*}
The quantities relevant for the evaluation of the correlation functions are $F^{(m,1)}$.
To see how we can build these quantities, we note that $F^{(m+1,n)}_k = \sum_{k'} F^{(m,n)}_{k'} \left( \mat{U}_B \right)_{k',k}(\Delta t)$.
Thus it suffices to begin our discussion with
\begin{align*}
  F^{(1,n)}_k &= \sum_{k'} U_{0, k'}(n \Delta t) \left( \mat{U}_B \right)_{k',k}(\Delta t) \equiv \sum_{k'} F^{(0,n)}_{k'} \left( \mat{U}_B \right)_{k',k}(\Delta t) \\
              &= U_{0,k}\Big((n+1)\Delta t\Big) - F(n \Delta t) U_{0, k}(\Delta t) \\
  &\equiv -F(n\Delta t) F^{(0,1)}_k + F^{(0,n+1)}_k.
\end{align*}
Observe that this can be written in the form
\begin{align*}
F^{(1,n)}_k &= 
\begin{pmatrix}
  - F(1\Delta t) \mat{I} & \mat{I} & 0 & 0 & \cdots & \\
  - F(2\Delta t) \mat{I} & 0 & \mat{I} & 0 &  & \\
  - F(3\Delta t) \mat{I} & 0 & 0 & \mat{I} &  & \\
  \vdots &  &  &  & \ddots &
\end{pmatrix}
\begin{pmatrix}
  \vec{U}_{0,k}(1\Delta t) \\
  \vec{U}_{0,k}(2\Delta t) \\
  \vec{U}_{0,k}(3\Delta t) \\
  \vdots
\end{pmatrix}.
\end{align*}
Importantly, this relation is recursive, so that
\begin{align*}
  F^{(m+1,n)}_k &= -F(n\Delta t) F^{(m,1)}_k + F^{(m,n+1)}_k \\
  \Longrightarrow F^{(m+1,n)}_k &= 
\begin{pmatrix}
  - F(1\Delta t) \mat{I} & \mat{I} & 0 & 0 & \cdots & \\
  - F(2\Delta t) \mat{I} & 0 & \mat{I} & 0 &  & \\
  - F(3\Delta t) \mat{I} & 0 & 0 & \mat{I} &  & \\
  \vdots &  &  &  & \ddots &
\end{pmatrix}^{m+1}
\begin{pmatrix}
  \vec{U}_{0,k}(1\Delta t) \\
  \vec{U}_{0,k}(2\Delta t) \\
  \vec{U}_{0,k}(3\Delta t) \\
  \vdots
\end{pmatrix}.
\end{align*}
Finally, we can recover the quantities of interest $F^{(m,1)}_k$ by projecting on to the $n=1$ subspace
\begin{align}
\sum_j U_{0,j} \left( \mat{U}_B^m \right)_{j,k} \equiv F^{(m,1)}_k &= \begin{pmatrix} \mat{I} & 0 & 0 & \cdots \end{pmatrix} 
\begin{pmatrix}
  - F(1\Delta t) \mat{I} & \mat{I} & 0 & 0 & \cdots & \\
  - F(2\Delta t) \mat{I} & 0 & \mat{I} & 0 &  & \\
  - F(3\Delta t) \mat{I} & 0 & 0 & \mat{I} &  & \\
  \vdots &  &  &  & \ddots &
\end{pmatrix}^{m}
\begin{pmatrix}
  \vec{U}_{0,k}(1\Delta t) \\
  \vec{U}_{0,k}(2\Delta t) \\
  \vec{U}_{0,k}(3\Delta t) \\
  \vdots
\end{pmatrix}.
\end{align}

Furthermore, the matrix $\mat{I}_B^{(n)}$ can similarly be shown to equal
\begin{align}
\left[ \mat{I}_B^{(n)} \right]_{k,k'} &= \delta_{k,k'} - \sum\limits_{r=0}^{n-1}  \Bigg[ \left( \mat{U}_B^{\dagger} \right)^r_{k,q} \Big( \mat{U}(\Delta t)^{\dagger} \Big)_{q,0} \Bigg] \Bigg[ \Big( \mat{U}(\Delta t) \Big)_{0,q'} \left( \mat{U}_B \right)^r_{q',k'} \Bigg].
\end{align}
Remarkably, it only deviates from the identity by a sum of outer products, i.e., $\mat{I}_B^{(n)} = \mat{I} - \sum_r |\vec{v}^{(r)}\rangle\langle \vec{v}^{(r)}|$.
This allows for a straightforward inversion, if we first define the adjacency matrix $\mat{A}$,
\begin{align}
\mat{A}_{n} &= 
\begin{pmatrix}
  \langle \vec{v}^{(0)} | \vec{v}^{(0)} \rangle & \cdots & \langle \vec{v}^{(0)} | \vec{v}^{(n-1)} \rangle \\
  \vdots & \ddots & \vdots \\
  \langle \vec{v}^{(n-1)} | \vec{v}^{(0)} \rangle & \cdots & \langle \vec{v}^{(n-1)} | \vec{v}^{(n-1)} \rangle
\end{pmatrix}.
\end{align}
The inverse is simply expressed as
\begin{align*}
\Big( \mat{I} - \sum_{r=0}^{n-1}|\vec{v}^{(r)}\rangle\langle \vec{v}^{(r)}| \Big)^{-1} &= \mat{I} + \sum_{i,j=0}^{n-1} |\vec{v}^{(i)}\rangle \left( \frac{1}{\mat{I} - \mat{A}_n} \right)_{ij} \langle \vec{v}^{(j)} |.
\end{align*}
For notational convenience, the vector $|\vec{v}^{(n)}\rangle$ is defined such that its the $k$-th entry is
\begin{align}
|\vec{v}^{(n)}\rangle_k &= \sum\limits_q \left( \mat{U}_B^{\dagger} \right)^n_{k,q} \Big( \mat{U}(\Delta t)^{\dagger} \Big)_{q,0}.
\end{align}
Thus, we can directly compute the $\left[ \mat{I}_B^{(m)} \right]^{-1}$ appearing in the correlation function $G^{++}_{<}(m,n)$.

In a similar way, we can compute the normalization factor $Z_N$ in terms of $Z = \operatorname{det}(\mat{I} + e^{-\beta \mat{h}_B})$ by defining a different adjacency matrix $\mat{B}_N$,
\begin{align}
\mat{B}_{N} &= 
\begin{pmatrix}
  \langle \vec{v}^{(0)} | (1 + e^{\beta \mat{h}_B} )^{-1} | \vec{v}^{(0)} \rangle & \cdots & \langle \vec{v}^{(0)} | (1 + e^{\beta \mat{h}_B} )^{-1} | \vec{v}^{(N-1)} \rangle \\
  \vdots & \ddots & \vdots \\
  \langle \vec{v}^{(N-1)} | (1 + e^{\beta \mat{h}_B} )^{-1} | \vec{v}^{(0)} \rangle & \cdots & \langle \vec{v}^{(N-1)} | (1 + e^{\beta \mat{h}_B} )^{-1} | \vec{v}^{(N-1)} \rangle
\end{pmatrix}.
\end{align}
The normalization is then given by
\begin{align}
Z_N &= Z \operatorname{det}(\mat{I} - \mat{B}_N).
\end{align}

The ability to compute $\mat{I}_B^{(n)}$ and its inverse immediately allows us to evaluate the bath traces and yields the quantities
\begin{align}
  G^{++}_{>}(m,n) &= - F(\Delta t) \sum\limits_{k=0}^{n-1} \left( \mat{I} - \frac{1}{\mat{I} - \mat{B}_N} (\mat{I} - \mat{A}_N) \right)_{m,k} \left( \frac{1}{\mat{I} - \mat{A}_n} \right)_{k,n-1}, \\
  G^{++}_{<}(m,n) &= - F(\Delta t) \left[ -\delta_{m-1,n} + \left( \frac{1}{\mat{I} - \mat{A}_m} \right)_{n,m-1} \right] - G^{++}_{>}(n,m) \nonumber \\
  &= F(\Delta t) \delta_{n,m-1} - F(\Delta t) \sum\limits_{k=0}^{m-1} \left( \frac{1}{\mat{I} - \mat{B}_N} (\mat{I} - \mat{A}_N) \right)_{n,k} \left( \frac{1}{\mat{I} - \mat{A}_m} \right)_{k,m-1}, \\
  G_{+}(m,n) &= -\delta_{m,n} + \left( \frac{1}{\mat{I} - \mat{B}_N} \right)_{n,m}, \\
\begin{split}
  G_{-}(m,n) &= - |F(\Delta t)|^2 \Bigg[ - \delta_{m,n} \left( \frac{1}{\mat{I} - \mat{A}_m} \right)_{m-1,m-1} \\
    &\qquad + \sum\limits_{k=0}^{m-1} \sum\limits_{\ell=0}^{n-1} \left( \frac{1}{\mat{I} - \mat{A}_m} \right)_{m-1,k} \left( (\mat{I} - \mat{A}_N) \frac{1}{\mat{I} - \mat{B}_N} (\mat{I} - \mat{A}_N) \right)_{k,\ell} \left( \frac{1}{\mat{I} - \mat{A}_n} \right)_{\ell,n-1} \Bigg].
\end{split}
\end{align}

\section{\label{sec:SM-berezin-integration}Computations with the MPS-IF}

\begin{figure}[h]
\centerline{\includegraphics[scale=0.7]{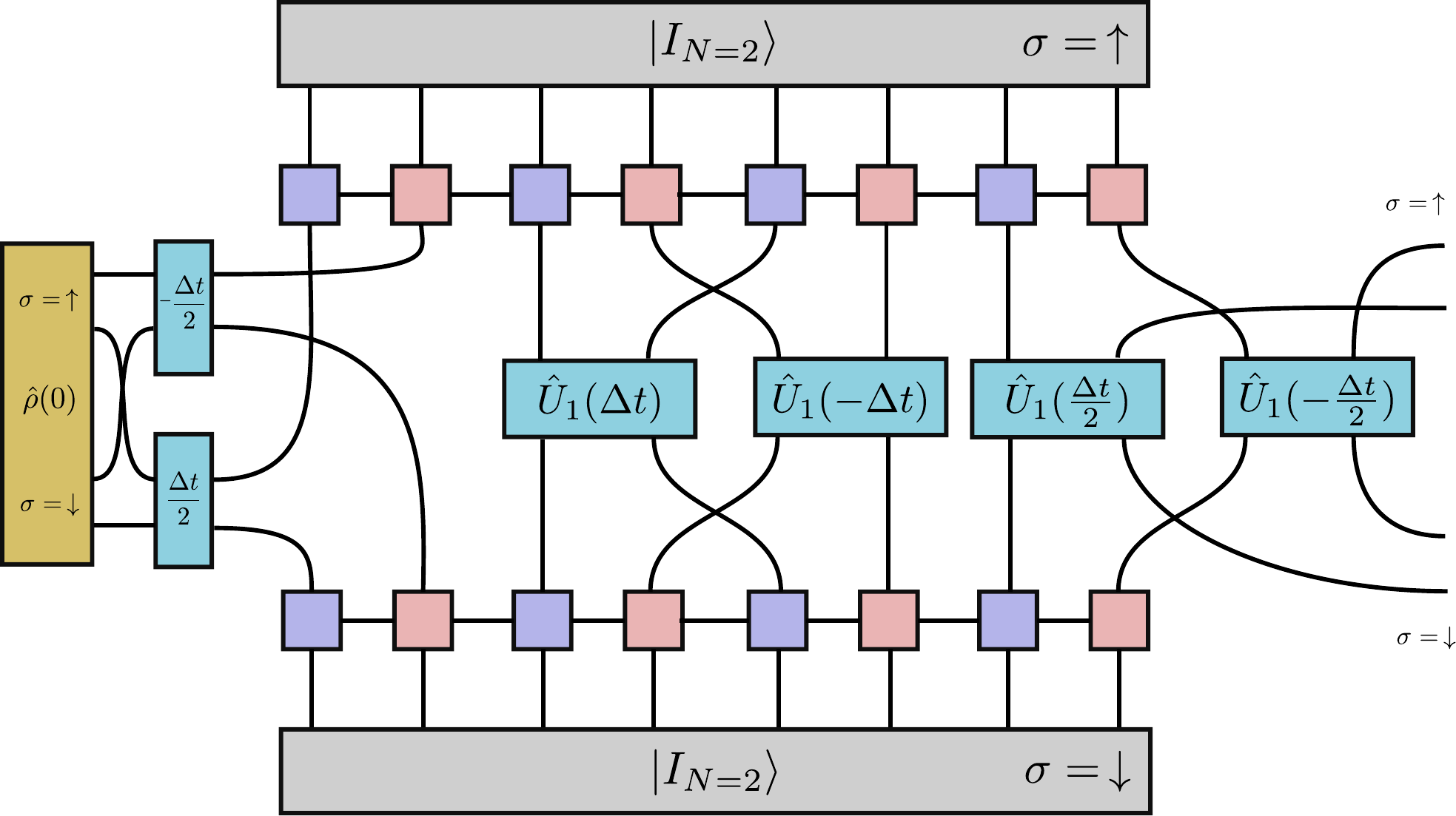}}
\caption[]{\label{fig:c-basis-SIAM} Schematic depiction of the tensor network used construct the impurity state $\op{\rho}(2\Delta t)$ for the single impurity Anderson model, from the MPS-IF.}
\end{figure}

In order to turn the influence functional into an object that can be contracted with the pure system propagators $\op{U}_0$, we must first convert from the fermionic representation back to the computational basis, \nnref{eq:unsplit-IF}{Eq.~(}{)}.
This can be done straightforwardly from the Grassmann integral over the $\eta$ variables. 
In terms of the mapping from Grassmann variables to fermions, we note that the equivalent of Grassmann integration is
\begin{align*}
\int d\xi_n \cdots d\xi_1 \, (\cdots) \quad \Longrightarrow \quad \langle \vec{0}| \op{c}_n \cdots \op{c}_1 \, (\cdots).
\end{align*}
Finally, we must choose an ordering of the ``fermions'' on the temporal lattice.
While this is mostly arbitrary, the time-locality of the correlations suggests that we should keep neighboring time points close together on the temporal lattice.
Hereafter, we take for $n = 0, \ldots, N-1$,
\begin{align}
\begin{split}
  \gF{\eta}{n^+} &\Longleftrightarrow \op{c}_{4n+1}^{\dagger}, \\
  \gBc{\eta}{n^+} &\Longleftrightarrow \op{c}_{4n+2}^{\dagger}, \\
  \gFc{\eta}{n+1^-} &\Longleftrightarrow \op{c}_{4n+3}^{\dagger}, \\
  \gB{\eta}{n+1^-} &\Longleftrightarrow \op{c}_{4n+4}^{\dagger}.
\end{split}  
\end{align}
Combining this with \nnref{eq:c-basis-kernels}{Eq.~(}{)}, along with the overlaps stemming from the resolution of the identity, we find that the influence functional can be easily transformed into the computational basis by introducing two tensors, one for variables on the forward contour and another for the backwards.
The basis change is facilitated by the red (backward contour) and blue (forward contour) tensors in \nnref{fig:c-basis-SIAM}{Fig.~}{}, which are connected by an auxiliary bond of dimension 2.
This function of this bond is to accumulate sign changes incurred by the presence of fermions on preceding (leftwards) sites.
In \nnref{fig:c-basis-SIAM}{Fig.~}{} we show a schematic diagram of how the impurity state $\op{\rho}(n\Delta t)$ can be extracted from the MPS-IF representing $|I_n\rangle$ using these basis change tensors.
Note that when the bath fermions of spin-up and spin-down are equivalent, their influence functionals (the upper and lower MPSs in gray) are identical.

\end{document}